\documentclass[preprint,prc,a4paper,tightenlines,nofootinbib,superscriptaddress]{revtex4}
\usepackage{amssymb}
\usepackage{amsmath}
\usepackage{graphicx}
\usepackage{bm} 

\usepackage[active]{srcltx}

\newcommand{\zx}{\taux}
\newcommand{\zm}{\taum}
\newcommand{\zp}{\taup}
\newcommand{\pp}{\phantom{+}} 
\newcommand{\mystructa}{S_1 \cdot l}
\newcommand{\mystructb}{\sdotltild }
\newcommand{\mystructc}{\sdotl \, \frac{\vdotltild}{2\mN}}
\newcommand{\mystructd}{\sdotltild \, \frac{\vdotl}{2\mN}}
\newcommand{\mystructe}{S_1 \cdot (\pone {+} \ponepr) \,\frac{\vdotl}{2\mN}}
\newcommand{\mystructf}{S_1 \cdot (\pone {+} \ponepr) \,\frac{\vdotltild}{2\mN}}
\newcommand{\mystructg}{\sdotl \frac{2 \vdotq}{-\vdotl + i \nolik}}
\newcommand{\mystructh}{\sdotltild \frac{2 \vdotq}{-\vdotl + i \nolik}}
\newcommand{\mystructi}{S_1 \cdot k_1 \frac{ q \cdot (l+\ltild)}{k_1^2-\mpi^2+i\nolik}}
\newcommand{\taup}{\tau_{\!+}}
\newcommand{\taum}{\tau_{\!-}}
\newcommand{\taux}{\tau_{\!\times}}

\newcommand{\maM}{\mathcal{M}}

\newcommand{\boldpi}{\bm{\pi}}
\newcommand{\boldtau}{\bm{\tau}}
\newcommand{\boldphi}{\bm{\phi}}
\newcommand{\be}{\begin{eqnarray}}
\newcommand{\ee}{\end{eqnarray}}
\newcommand{\beq}{\begin{equation}}
\newcommand{\eeq}{\end{equation}}

\newcommand{\mpi}{\ensuremath{m_\pi}}   
\newcommand{\mN}{\ensuremath{m_{\!N}}}   
\newcommand{\gA}{\ensuremath{g_{\!A}}} 
\newcommand{\fpi}{\ensuremath{f_{\!\pi}}} 
\newcommand{\nolik}{\ensuremath{0}}    
\newcommand{\Bnotation}{\ensuremath{B_2}}  
\newcommand{\Dnotation}{\ensuremath{D_2}}
\newcommand{\intdl}{\int \! \! \! \frac{d^{4}l}{(2 \pi)^4} \,}   
\newcommand{\NLO}{NLO}
\newcommand{\NNLO}{N$^2$LO} 
\newcommand{\NNNLO}{N$^3$LO} 
\newcommand{\NNNNLO}{N$^4$LO}

\newcommand{\pone}{\ensuremath{p_{1}}}
\newcommand{\ponezero}{\ensuremath{p_{10}}}
\newcommand{\ponevec}{\ensuremath{\vec p_{1}}}
\newcommand{\ponekin}{\ensuremath{\frac{\ponevec{}^{\!\!2}}{2 \mN}}}

\newcommand{\ptwo}{\ensuremath{p_{2}}}

\newcommand{\ponepr}{\ensuremath{p_{1}^{\,\prime}}}
\newcommand{\ponezeropr}{\ensuremath{p_{10}^{\,\prime}}}
\newcommand{\ponevecpr}{\ensuremath{\vec p_{1}^{\,\prime}}}

\newcommand{\ptwopr}{\ensuremath{p_{2}^{\,\prime}}}

\newcommand{\qzero}{\ensuremath{q_{0}}}
\newcommand{\qvec}{\ensuremath{\vec q}}

\newcommand{\lzero}{\ensuremath{l_{0}}}
\newcommand{\lvec}{\ensuremath{\vec l}}

\newcommand{\ltild}{\ensuremath{\tilde{l}}}
\newcommand{\ltildzero}{\ensuremath{\tilde{l}_{0}}}
\newcommand{\ltildvec}{\ensuremath{\vec{\tilde{l}}}}

\newcommand{\vdotl}{\ensuremath{v \cdot l }}
\newcommand{\vdotltild}{\ensuremath{v \cdot \ltild }}
\newcommand{\vdotk}{\ensuremath{v \cdot k_1 }}
\newcommand{\vdotq}{\ensuremath{v \cdot q }}

\newcommand{\sdotl}{\ensuremath{S_1 \cdot l }}
\newcommand{\sdotltild}{\ensuremath{S_1 \cdot \ltild }}

\newcommand{\sdotk}{\ensuremath{S_1 \cdot k_1 }}

\newcommand{\sdotlmn}{\ensuremath{\frac{S_1 \cdot l }{2\mN}}}
\newcommand{\sdotltildmn}{\ensuremath{\frac{S_1 \cdot \ltild }{2\mN} }}

\newcommand{\sdotpppprmn}{\ensuremath{\frac{S_1 \cdot (\pone + \ponepr) }{2\mN} }}

\begin{document}

\title{Pion production in nucleon-nucleon collisions in chiral
  effective field theory: next-to-next-to-leading order contributions}

\author{A.~A.~Filin}
\affiliation{\footnotesize Institut f\"ur Theoretische Physik II, Ruhr-Universit\"at Bochum, D-44780 Bochum, Germany}
\affiliation{\footnotesize Institute for Theoretical and Experimental Physics,
 117218, B.~Cheremushkinskaya 25, Moscow, Russia}

\author{V.~Baru}
\affiliation{\footnotesize Institut f\"ur Theoretische Physik II, Ruhr-Universit\"at Bochum, D-44780 Bochum, Germany}
\affiliation{\footnotesize Institute for Theoretical and Experimental Physics,
 117218, B.~Cheremushkinskaya 25, Moscow, Russia}

\author{E.~Epelbaum }
\affiliation{\footnotesize Institut f\"ur Theoretische Physik II, Ruhr-Universit\"at Bochum, D-44780 Bochum, Germany}

\author{C.~Hanhart}
\affiliation{\footnotesize Institut f\"{u}r Kernphysik,  (Theorie) and J\"ulich Center for Hadron Physics,
 Forschungszentrum J\"ulich,  D-52425 J\"{u}lich, Germany and\\
Institute for Advanced Simulation, Forschungszentrum J\"ulich,  D-52425 J\"{u}lich, Germany}

\author{H.~Krebs}
\affiliation{\footnotesize Institut f\"ur Theoretische Physik II, Ruhr-Universit\"at Bochum, D-44780 Bochum, Germany}

\author{A.~E.~Kudryavtsev}
\affiliation{\footnotesize Institute for Theoretical and Experimental Physics,
 117218, B.~Cheremushkinskaya 25, Moscow, Russia}

\author{F.~Myhrer}
\affiliation{\footnotesize Department of Physics and Astronomy, University of South Carolina,
Columbia, SC 29208, USA}

\begin{abstract}
A complete calculation of the pion-nucleon loops that contribute to
the transition operator for $NN\to NN\pi$ up-to-and-including next-to-next-to-leading
order (N$^2$LO)  in chiral effective field theory
near threshold is presented. The evaluation is  based on  the so-called momentum counting scheme, which 
takes into account the relatively large momentum of the initial nucleons 
inherent in pion-production reactions. 
We show that the significant
cancellations between the loops found  at  next-to-leading order (NLO)
in the earlier studies are also operative at N$^2$LO. 
In particular,
the $1/m_N$ corrections (with $m_N$ being the nucleon mass) to loop diagrams 
cancel at  N$^2$LO, as do  
the contributions  of the pion loops involving the low-energy constants 
$c_i$, $i=1\ldots 4$. 
In contrast to the NLO calculation however, the
cancellation   of loops at N$^2$LO is incomplete, yielding a non-vanishing
contribution to the transition amplitude. 
Together with the
one-pion exchange tree-level operators, the loop contributions
provide the long-range part of the production operator.
Finally, we discuss the phenomenological implications of these findings. 
In particular, we
find that the amplitudes generated by the N$^2$LO pion loops yield contributions comparable in
size with the most important phenomenological 
heavy-meson exchange amplitudes. 
\end{abstract}

\maketitle

\section{Introduction}

The reaction $NN\to NN\pi$ has been extensively studied both
theoretically and experimentally over the past decades.
However, the near-threshold regime is still not yet fully
understood.
After the first high-quality data for $pp\to pp\pi^0$~\cite{IUCF1}
became available  
(further experimental data can be found in, e.g., the  review article \cite{hanhart04}, 
with the latest measurements in Refs.~\cite{ANKEpi0,ANKEpi-}), 
it quickly became clear that the original models failed to 
reproduce the new data.
 For example, 
the model of
Ref.~\cite{koltunundreitan} fell short by a factor of two for the reaction
$pp\to d\pi^+$ and by an order of magnitude for  $pp\to pp\pi^0$. 
Various attempts were
made to identify the phenomenological mechanisms responsible for this
discrepancy. 

The first theoretical paper to explain quantitatively the cross section $pp\to
pp\pi^0$ was Ref.~\cite{Lee}. The new contribution in \cite{Lee} originated
from the short ranged, irreducible currents constructed directly from the
nucleon-nucleon potential. 
A phenomenological interpretation of this
mechanism was provided in Ref.~\cite{HGM}, where the  exchange of heavy
mesons (mostly $\sigma$ and $\omega$) followed by a pion emission via a
nucleon-antinucleon pair (the so-called z-mechanism)
was calculated. The mechanism was also shown to
provide the missing strength for $pp\to d\pi^+$ in
Refs.~\cite{Hpipl,jounicomment}.
An alternative mechanism is based on the pion-nucleon 
rescattering diagram where the off-shell pion-nucleon amplitude 
plays a crucial role. 
It is well-known that the isoscalar
pion nucleon scattering length is very small --- see Refs.~\cite{ourpid} for
its most recent determination --- as a result of a cancellation of
individually sizable terms which have different energy dependences. 
It therefore appeared 
natural that in the off-shell kinematics relevant for the pion production
reaction the amplitudes are significantly enhanced. 
This mechanism
was also shown to be capable of describing the experimental data in both $pp\to
pp\pi^0$ \cite{eulogio,unsers}
as well as $pp\to d\pi^+$~\cite{unserd} reactions. At this point there was
no way to decide which of the mechanisms described captures the correct
physics.

Since pion interactions are largely controlled by the chiral symmetry of the
strong interaction, one might naturally expect that chiral
perturbation theory (ChPT)  provides the proper tool to resolve the
above mentioned discrepancy.
However, the use of the standard ChPT power counting, which is based on the
assumption that all relevant momenta are effectively of the order of
the pion mass, was not very successful. 
The first calculations in this
framework were done at  tree level up to \NNLO{}  for 
both $pp\to pp\pi^0$~\cite{cohen,park,sato}
as well as for $pp\to d\pi^+$~\cite{rocha,unserd}.
These studies revealed, in particular,  that the discrepancy between theory
and experiment increases  for the neutral channel due to a destructive interference of
the direct pion production and the isoscalar 
rescattering contributions at  NLO in standard counting.
In addition, some loop contributions at \NNLO{}  were found in Refs.~\cite{DKMS,Ando} to
be larger than the NLO contribution,   
revealing a 
problem regarding the convergence of the standard ChPT power counting. 

It was soon realized that the large initial nucleon
momentum at threshold $p$, $p =|\vec p \,|= \sqrt{m_N\mpi} $,
which is significantly larger than the pion mass $\mpi$,
requires the modification of the standard power counting.
The corresponding expansion parameter in the new scheme is
\begin{equation}
\chi= p/\Lambda_\chi \simeq  \ 0.4,
\label{eq:expanpar}
\end{equation}
with $\Lambda_\chi$ being the chiral symmetry breaking scale of the order of 1 GeV.
Here and in what follows, this power counting will be referred to as
the momentum counting scheme (MCS). This modification was proposed in
Refs.~\cite{cohen,rocha} while the proper way to treat this scale was
first presented in Ref.~\cite{ch3body}
and  implemented in Ref.~\cite{HanKai}, see Ref.~\cite{hanhart04} for
a review article.
The MCS expansion is performed with two distinct parameters,
namely the initial nucleon momentum $p$ and the pion mass $\mpi$, where 
$\mpi/p \sim p/\Lambda_\chi$.
The pion loop diagrams  
start to contribute at a given order in the expansion parameter, which 
can be identified based on the power counting, and, unlike the standard ChPT power counting, 
continue to contribute at all higher MCS orders. 

Due to the fact that the Delta-nucleon ($\Delta$-N) mass splitting
is numerically of the order of $p$, the Delta-isobar should be 
explicitly included as a dynamical degree of freedom~\cite{cohen}. 
This
general argument was confirmed numerically in phenomenological
calculations~\cite{jouni,ourdelta,ourpols}, 
see also Refs.~\cite{cohen,rocha,HanKai,NNpiMenu} where the effect
of the $\Delta$ in $NN\to NN\pi$ was studied within chiral EFT.
However, in this paper we focus on contributions from nucleons
and pions only. The $\Delta$ degree of freedom will be included in a
subsequent publication.

In the MCS, pion p-waves are given by tree level diagrams up to \NNLO{}
and the corresponding calculations of Refs.~\cite{ch3body,newpwave}
showed a satisfactory agreement with the data. 
Meanwhile, for pion s-waves loop diagrams 
start to  contribute individually already at NLO. However,
they turned out to cancel completely 
both for the neutral~\cite{HanKai} and charged~\cite{lensky2}
pion production, a result which is reproduced
in this paper.  
To obtain this result for charged pion production,
it is crucial to consistently take into account
a contribution related to nucleon recoil in the $\pi N$ vertex
as explained in detail in Ref.~\cite{lensky2}.
As a by-product of the consistent treatment of nucleon recoil effects
in Ref.~\cite{lensky2}, the rescattering one-pion exchange amplitude at LO was
found to be enhanced by a factor of $4/3$
which was sufficient to overcome the apparent discrepancy 
with the data in the charged channel.
The first attempts to study the subleading loop contributions were 
taken in  Refs.~\cite{subloops,ksmk09}.

In this paper we advance the analysis for $NN\to NN\pi$ at threshold to \NNLO{}.
In particular, we evaluate all loop contributions at \NNLO{} 
that involve pion and nucleon degrees
of freedom. A complete calculation of all operators at \NNLO{} 
(tree level and loops) including
the $\Delta$ degree of freedom, and the subsequent  convolution with the
pertinent $NN$ interactions in the initial and final states  will be
reported elsewhere.
We will show in this paper that also at \NNLO{} significant cancellations
occur and that only very few loop topologies contribute to the final amplitude.

The paper is structured as follows.
In Sec.~\ref{sec:formalism} we present our formalism and discuss the hierarchy of diagrams as
follows from our power counting.
The next two sections are devoted to a detailed discussion of 
the results for the loop topologies proportional
to the axial-vector nucleon coupling constant $\gA$ to the third (Sec.~\ref{sec:calcga3})
and the first power (Sec.~\ref{sec:calcga1}).
In particular, we reproduce the cancellation of all NLO terms found in 
Ref.~\cite{lensky2} and demonstrate
that a similar cancellation pattern also takes place among the 
loop contributions at \NNLO{}. 
In the latter
case, however, the  cancellation is not  complete.
Sec.~\ref{sec:sumTPE} contains a compact summary of the results of 
Secs.~\ref{sec:calcga3} and \ref{sec:calcga1}.
Here, we also give explicitly the finite loop contributions which
survive the above mentioned cancellation. 
In Sec.~\ref{sec:reg} the regularization procedure
for the loop integrals is outlined. 
In this section we also compare the finite pieces of our loops at \NNLO{} to the
size of the contact $4N\pi$ operator estimated based on phenomenological calculations.
Finally, in Sec.~\ref{sec:outlook} we summarize the results of the paper
and discuss phenomenological implications of the observed cancellation
of loop contributions.

\section{Formalism and power counting}
\label{sec:formalism}

\subsection{Reaction amplitude and Lagrangian densities}
\label{sec:ampl}

The most general form of the threshold amplitude for
the pion production reaction $N_1(\vec p\,)+N_2(-\vec p\,) \to N+N +\pi$ in
the center-of-mass frame, can be written as:
\begin{eqnarray}
M_{th}(NN\to NN \pi) &=&  \, \Big[ {\cal A}_1\, i\,(\vec \sigma_1 -\, \vec \sigma_2) \cdot \vec p
+{\cal A}_2\,(\vec \sigma_1 \times \vec \sigma_2)\cdot \vec p \Big]
\,\,\, (\boldtau_1+\boldtau_2)\cdot \boldphi^{\,*} \nonumber \\ & & +
\,(\vec \sigma_1 + \vec \sigma_2)\cdot \vec p  \,\,\, \Big[ {\cal B}_1 \,i\,
(\boldtau_1- \,\boldtau_2) + {\cal B}_2\,(\boldtau_1 \times \boldtau_2)\Big] \cdot
\boldphi^{\,*} \,,
\label{eq:Mthrred}
\end{eqnarray}
where $\vec\sigma_{1,2}$ and $\boldtau_{1,2}$ are the spin and isospin operators of
nucleons 1 and 2. The final state pion's three-component isospin wave function
is denoted by $\boldphi$, e.g. $\boldphi= (0,0,1)$
for $\pi^0$-production and $\boldphi = (1,i,0)/\sqrt2$ for $\pi^+$-production.

However, as follows from the angular momentum conservation and the Pauli selection  rule for the
$NN$ system, a final s-wave pion in $NN\to NN\pi $ can be produced via two
angular momentum
transition channels only, namely $^3\!P_0\to {^1\!}S_0s$  and   $^3\!P_1\to {^3\!}S_1s$,
where we use the spectroscopic notation $^{2S+1}L_J$ for the $NN$ states  while 
the lower case $s$  corresponds to the $l=0$ pion partial wave in the overall  cms. 
Therefore, the two spin-isospin structures in Eq.~\eqref{eq:Mthrred} are redundant, and
the reaction amplitude, that acknowledges the Pauli principle,
can be  rewritten, without loss of generality,
as  \cite{novel,HanKai}
\begin{eqnarray}
M_{th}(NN\to NN \pi) =  \, {\cal A}\,(\vec \sigma_1 \times \vec \sigma_2)\cdot \vec p
\,\,\, (\boldtau_1+\boldtau_2)\cdot \boldphi^{\,*} 
+{\cal B}\,
\,(\vec \sigma_1 + \vec \sigma_2)\cdot \vec p  \,\,\, (\boldtau_1 \times \boldtau_2)\cdot
\boldphi^{\,*} \,,
\label{eq:Mthr}
\end{eqnarray}
with  ${\cal A}= {\cal A}_1 +{\cal A}_2$ and ${\cal B}= {\cal B}_1 +{\cal B}_2$.
To derive  \eqref{eq:Mthr}  
we used the fact that the spin (isospin) matrix element for the operator
$\hat O=i(\vec \sigma_1 - \vec \sigma_2)\cdot \vec p$ \
$( \hat O=i\, (\boldtau_1- \,\boldtau_2)\cdot \boldphi^{\,*})$
is equal to that of
$\hat O=(\vec \sigma_1 \times \vec \sigma_2)\cdot \vec p$ \
$(\hat O=\,(\boldtau_1 \times \boldtau_2)\cdot \boldphi^{\,*})$  for s-wave pion production.

The amplitude ${\cal A}$ in Eq.~(\ref{eq:Mthr})  contributes to  $^3\!P_0\to {^1\!}S_0s$,
which  is the relevant transition  amplitude for neutral pion production in $pp \to pp\pi^0$.
Conversely, the  amplitude ${\cal B}$ in Eq.~(\ref{eq:Mthr})  contributes to
the charge pion production in ${pp\to d\pi^+}$,
driven by  the $^3\!P_1\to {^3\!}S_1s$ transition operator.
Furthermore, in some channels such as e.g. $pp\to pn\pi^+$,
both amplitudes  ${\cal A}$  and  ${\cal B}$ contribute
in a certain linear combination.

It is  convenient to write down the  threshold reaction amplitudes in the form
where the relevant spin-angular structure of the initial and final nucleon pairs
are shown explicitly \footnote{The
connection of the amplitudes ${\cal A}$ and ${\cal B}$  to the observables is given in,
e.g.,  Ref.~\cite{newpwave} }
\begin{eqnarray}\nonumber
\maM_{pp\to pp\pi^0} &=&4i {\cal A} (\vec {\mathcal{S}}\cdot \hat p)\mathcal{I}^{\prime\dagger},\\
\maM_{pp\to d\pi^+}&=&-2\sqrt{2}i {\cal B}\,(\vec{\mathcal{S}}\times \hat p\,)\cdot \vec\varepsilon\ .
\label{eq:ampproj}
\end{eqnarray}
 Here $\vec\varepsilon$ is the deuteron polarization vector,
$\hat p$ is the unit vector of the initial  relative momenta of two  nucleons, and
$\vec {\mathcal{S}}=\chi^T_2{\sigma_y}\vec \sigma\chi_1/\sqrt{2}$, and
$\mathcal{I}^{\prime\dagger}=\chi^\dagger_{1^\prime}{\sigma_y}\chi^*_{2^\prime}/\sqrt{2}$
denote the normalized spin structures of the initial spin-triplet and
final spin-singlet states, respectively.

The main goal of this paper is to derive the contributions to $\cal A$ and $\cal B$
that originate from loop diagrams.
The loop diagrams can be separated in two different kinds:
the ones involving only pion and nucleon degrees of freedom
and the ones involving $\Delta$(1232) excitations in the intermediate
states.
In this paper we concentrate on the first kind only where we include all relevant contributions
at orders NLO and \NNLO{} in the MCS as detailed  in the next section.
An evaluation of all MCS \NNLO{} operators containing explicitly the $\Delta$ will
be presented in an upcoming publication.

Our calculations are based on the effective chiral Lagrangian in which  the
lowest-order (LO) $\pi N$ interaction terms read
in $\sigma$-gauge \cite{OvK,ulfbible}  
(more details on the pion-nucleon Lagrangian can be found, e.g., in Ref.~\cite{Fettes})
\begin{eqnarray}
 {\cal L}^{(1)}_{\pi\!N}  &=&
   N^{\dagger}\left[\frac{1}{4 f_{\pi}^{2}} \boldtau \cdot
         (\dot{\boldpi}\times{\boldpi})
         +\frac{g_{A}}{2 f_{\pi}}
         \boldtau\cdot\vec{\sigma}\left(\vec{\nabla}\boldpi
{+}\frac{1}{2f_\pi^2}\boldpi(\boldpi \cdot \vec \nabla \boldpi)
\right)
  \right]N +\cdots \ .
\label{eq:la0}
\end{eqnarray}
The next-higher order interaction terms have the form
\begin{eqnarray}\nonumber
 \hspace*{-0.2cm}{\cal L}^{(2)}_{\pi\!N}&=&
    \frac{1}{8\mN f_{\pi}^{2}}
    \bigg[ iN^{\dagger}\boldtau\cdot
        (\boldpi\times\vec{\nabla}\boldpi)\cdot\vec{\nabla}N + h.c.\bigg]\\
&{-}&\frac{g_{A}}{4 m_{N} f_{\pi}}\bigg[iN^{\dagger}\boldtau\cdot
\left(\dot{\boldpi}
{+}\frac{1}{2f_\pi^2}\boldpi(\boldpi \cdot  \dot{\boldpi})
\right)        \vec{\sigma}\cdot\vec{\nabla}N +  h.c.\bigg]
-\frac{\gA}{8 \mN \fpi^3} N^{\dagger} \boldpi \cdot (\vec{\sigma} \cdot \vec{\nabla})
        (\dot{\boldpi} \times \boldpi) N
\nonumber \\\nonumber &+&\frac{1}{f_{\pi}^{2}}N^{\dagger}\bigg[ \left(c_3+c_2-\frac{g_A^2}{8m_N}\right) \dot\boldpi^{2}
          - c_3 (\vec{\nabla}\boldpi)^{2}-2c_1\mpi^2 \boldpi^{2}\\
          &-&\frac{1}{2} \left(c_4 + \frac{1}{4m_{N}}\right)  \varepsilon_{ijk} \varepsilon_{abc}
          \sigma_{k} \tau_{c}
        \partial_{i}\pi_{a}\partial_{j}\pi_{b}\bigg] N    +\cdots  \, \ .
 \label{eq:la1}
\end{eqnarray}
\noindent
In the equations above $\fpi$ denotes the pion decay constant,
 $g_A$ is the axial-vector coupling of the nucleon, and
 $N$ ($\pi$)  corresponds to the nucleon (pion) field.
The ellipses represent further terms which are not relevant for the present
study.

The Lagrangian density for the leading 4$\pi$ vertex also needed for
the calculation reads in the $\sigma$-gauge:
\begin{eqnarray*}
 {\cal L}_{\pi\pi}^{(2)}  =
 \frac{1}{2\fpi^2} (\boldpi \cdot \partial^{\mu} \boldpi) (\boldpi \cdot \partial_{\mu} \boldpi )
 - \frac{\mpi^2}{8\fpi^2} \boldpi^4.
\end{eqnarray*}

The  loop diagrams (at \NNLO{}) lead to ultraviolet (UV) divergent
integrals. These UV divergencies are removed by expressing the bare
LECs accompanying the five-point contact vertex at the same order in
terms of renormalized ones. 
As a consequence of the Pauli principle,
only two independent linear combinations of Lagrangian contact terms contribute 
to the transition matrix elements ${\cal A}$ and ${\cal B}$.
We will denote  the corresponding amplitudes by
${\cal A}_\text{CT}$ and ${\cal B}_\text{CT}$ for future references.

\subsection{Diagrams and Power Counting}
\label{sec:PC}

In ChPT the expansion parameter is $Q/\Lambda_{\chi}$ where
$Q$ is identified either with a typical momentum of the process or
$\mpi$. The key assumption for convergence of the theory is $Q \ll \Lambda_{\chi}$.
As mentioned in the introduction, the reaction $NN\to NN\pi$ at
threshold involves
momenta of  ``intermediate range'' $p\approx \sqrt{\mpi \mN}$ larger  than $\mpi$ but still
smaller than the $\Lambda_{\chi}\sim \mN$.
In the MCS we are thus faced with a two-scale expansion.
For near threshold s-wave pion production,
the outgoing two-nucleon pair has a low relative three-momentum $p'$
and appears therefore  predominantly in S-wave.
We therefore assign  $p'$ an order $\mpi$ and introduce the expansion parameter
\begin{equation}
\chi\simeq\frac{p'}{p}\simeq\frac{\mpi}{p}\simeq\frac{p}{\mN}.
\label{expansionpapar}
\end{equation}
\begin{figure}[t]
\includegraphics{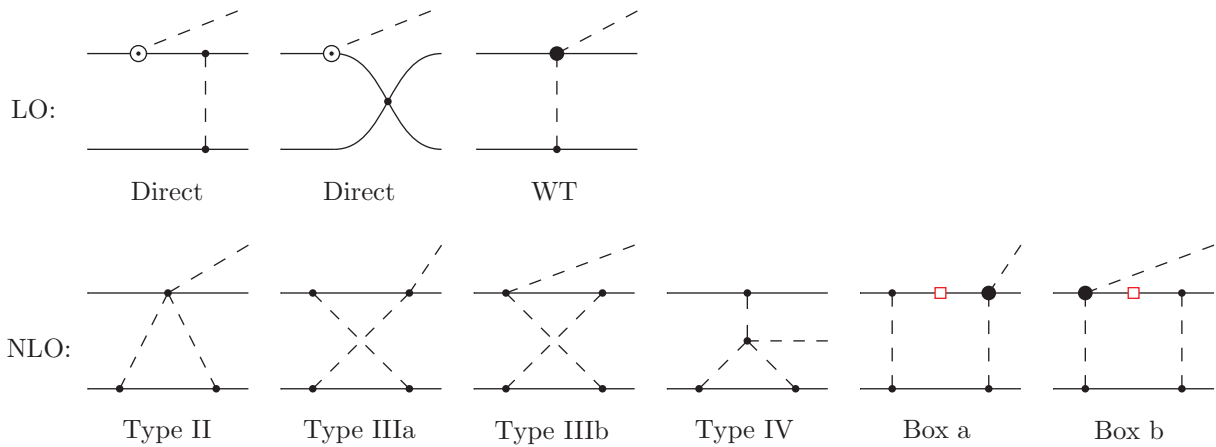}
\caption{\label{fig:allNLO}
Complete set of diagrams up to NLO (in the $\Delta$-less theory). Solid (dashed) lines denote nucleons (pions).
Solid dots correspond to the leading  vertices from
${\cal L}^{(1)}_{\pi\!N}$ and ${\cal L}_{\pi\pi}^{(2)} $, as given in the main text,
$\odot$ stands for the sub-leading vertices from  ${\cal L}^{(2)}_{\pi\!N}$
whereas the blob  indicates the possibility to have  both leading and subleading vertices.
The $NN$ contact interaction is represented by  the leading S-wave LECs  $C_S$ and $C_T$.
The red square in the box diagrams indicates 
that the corresponding nucleon propagator cancels with
parts of the $\pi N$ vertex and leads to the  irreducible contribution, 
see text for further details.
}
\end{figure}

The diagrams containing only pion and nucleon degrees of freedom
that contribute to the reaction $NN\to NN\pi$
up to NLO in our expansion, are shown in Fig.~\ref{fig:allNLO}.
Details of the  evaluations  of each of the loop diagrams can be found
in appendices \ref{sec:appga3} and \ref{sec:appga1}.
The first two diagrams in the first line are sometimes called the ``direct''
one-nucleon diagrams in the literature,
whereas the last (rightmost) diagram  is called the rescattering diagram.
We will discuss both next.

At leading  order  one needs to deal with  the
``direct'' pion emission from a single nucleon where
the nucleon recoil $\pi NN$ vertex
of  ${\cal L}^{(2)}_{\pi\!N}$ \eqref{eq:la1} is necessary in order to produce
an outgoing s-wave pion.
In addition, at LO there is a rescattering operator with
the Weinberg--Tomozawa (WT) $\pi\pi N N$ vertex
which, however, contributes only to the charged pion channel due to its  isovector nature.
In order to clarify the counting in MCS, we will concentrate on
the first of two   ``direct'' diagrams in Fig.~\ref{fig:allNLO}.
In this diagram each vertex attached to the pion propagator
involves a momentum $p$.
The pion propagator itself
involves a momentum $p$, i.e.~it is counted as $p^{-2}$,
whereas the  nucleon propagator only carries an energy $\propto \mpi$.
The $\mpi^{-1}$ of the  nucleon propagator cancels the
factor $\mpi$
of the s-wave pion production vertex, which counts as $\sim \mpi p/\mN$.
Thus, counting the ``momentum flow'' in the vertices and propagators of the diagram,
gives an order of magnitude estimate of the direct diagrams
(as well as the rescattering diagram) as
$p/\mN$.
These  diagrams are counted as LO in MCS.
Traditionally the LO direct diagrams have been evaluated numerically by including the
pion propagator in the  distorted $NN$ wave functions, i.e. only the
one-nucleon-pion production vertex gives the transition operator.
 Numerically, in the traditional distorted wave Born approximation approach,
the ``direct'' term appears to be significantly smaller than the estimate based
on our naive MCS's dimensional analysis.  This suppression comes from
two sources:
first,  there is the momentum mismatch between the initial and final
distorted nucleon wave functions~\cite{hanhart04} --- see also Ref.~\cite{BM}
for a more detailed discussion. Secondly, there are accidental
cancellations from the final state interaction present in both channels,
$pp\to pp\pi^0$ and $pp\to d\pi^+$, that are not accounted for in the power
counting. Specifically,
the $NN$ phase shift in the
$^1S_0$ partial wave relevant for $pp\to pp\pi^0$ crosses zero at an energy
close to the pion production threshold \cite{SAID}.  All realistic $NN$
scattering potentials that reproduce this feature show in the half-off-shell
amplitude at low energies a zero at off-shell momenta of a similar magnitude.
The exact position of the zero varies between different models, such that the direct
production amplitude turns out to be quite model dependent.
The suppression mechanism of the direct term for the reaction $pp\to d\pi^+$
comes from  a strong
cancellation between the deuteron S-wave and D-wave components.
Thus, it is not
surprising that numerically the ``direct'' terms in both channels
are about an order of magnitude smaller than the LO amplitude from the
rescattering diagram, which turns out to be consistent with the dimensional analysis.
Since this LO contribution is forbidden by selection rules for $pp\to
pp\pi^0$ while allowed for $pp\to d\pi^+$, one understands directly why
a theoretical understanding is a lot more difficult to achieve for the former
reaction.

\begin{figure}[t]
\includegraphics{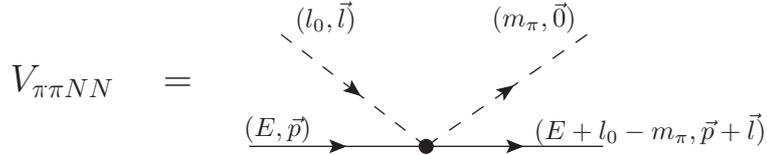}
\caption{\label{fig:vpipinn}
The $\pi N\to \pi N$ transition vertex:
definition of kinematic variables as used in Eq.~(\ref{eq:pipivert}).}
\end{figure}
At NLO, which corresponds to the order $p^2/\mN^2$,
loop diagrams illustrated in Fig.~\ref{fig:allNLO} start to contribute to the
s-wave pion production amplitude.
For the channel $pp\to pp\pi^0$ the sum of NLO diagrams type II, III and IV in
Fig.~\ref{fig:allNLO} is zero due to a cancellation between individual diagrams \cite{HanKai}.
However,  the same sum of diagrams II -- IV gives a finite answer for the channel
$pp\to d\pi^+$ \cite{HanKai}.
As a result the net contribution of these diagrams depends linearly on the $NN$
relative momentum which results in a large sensitivity to the
short-distance $NN$ wave functions~\cite{Gardestig}.
This puzzle was solved in Ref.~\cite{lensky2}, where it was demonstrated
that for the deuteron channel there is an additional contribution at NLO, namely the
box diagrams in Fig.~\ref{fig:allNLO}, stemming from the time-dependence of
the Weinberg--Tomozawa pion-nucleon  vertex.
To demonstrate this, we write the expression for the WT $\pi N\to \pi N$ vertex
in the notation of Fig.~\ref{fig:vpipinn}
as:
\begin{eqnarray}
V_{\pi\pi NN}&=&
l_0{+}\mpi {-}\frac{\vec l\cdot(2\vec p+\vec l)}{2\mN}
\nonumber \\
&=&
{2\mpi}+{\left(l_0{-}\mpi {+}E{-}\frac{(\vec l+\vec p)^2}{2\mN}+i\nolik \right)}-
{\left(E{-}\frac{\vec p\, ^2}{2\mN}+i\nolik\right)}
\ ,
\label{eq:pipivert}
\end{eqnarray}
where we kept the leading WT vertex and its nucleon recoil correction,
which are of the same order in the MCS, as explained below. 
For simplicity we omit the isospin dependence of the vertex.
The first term in the last line is the WT-vertex for  kinematics with the
on-shell incoming and outgoing nucleons,
the second term the
inverse of the outgoing nucleon propagator while the 
third one is the inverse of the
incoming nucleon propagator.
Note that for on-shell
incoming and outgoing nucleons, the expressions in brackets 
in \eqref{eq:pipivert} vanish, and
the $\pi N\to \pi N$ transition vertex
takes its on-shell value $2\mpi$ 
(even if the incoming pion is off-shell).
This is in contrast to standard phenomenological treatments~\cite{koltunundreitan},
where $l_0$ in the first line of  \eqref{eq:pipivert} is
identified with $\mpi/2$, the energy transfer in the on-shell kinematics for $NN\to NN\pi$,
but the recoil terms in Eq.~(\ref{eq:pipivert}) are not considered.
However, $\vec{p}^{\ 2}/\mN\approx \mpi$ so that
the recoil terms are to be kept in the vertices and  in
the   nucleon propagator\footnote{How to deal with the nucleon propagator 
in the MCS was shown in  Ref.~\cite{subloops}.}.
The MCS is explicitly designed to properly keep track of these recoil terms.
A second consequence of Eq.~(\ref{eq:pipivert}) is that  only the
first term leads to a reducible diagram  when the rescattering diagram with the
$\pi N\to \pi N$ vertex
is convoluted with $NN$ wave functions.
The second and third terms in Eq.~(\ref{eq:pipivert}), however,
lead to irreducible contributions, since one of the nucleon propagators is
cancelled. 
%
\begin{figure}[t]
\includegraphics{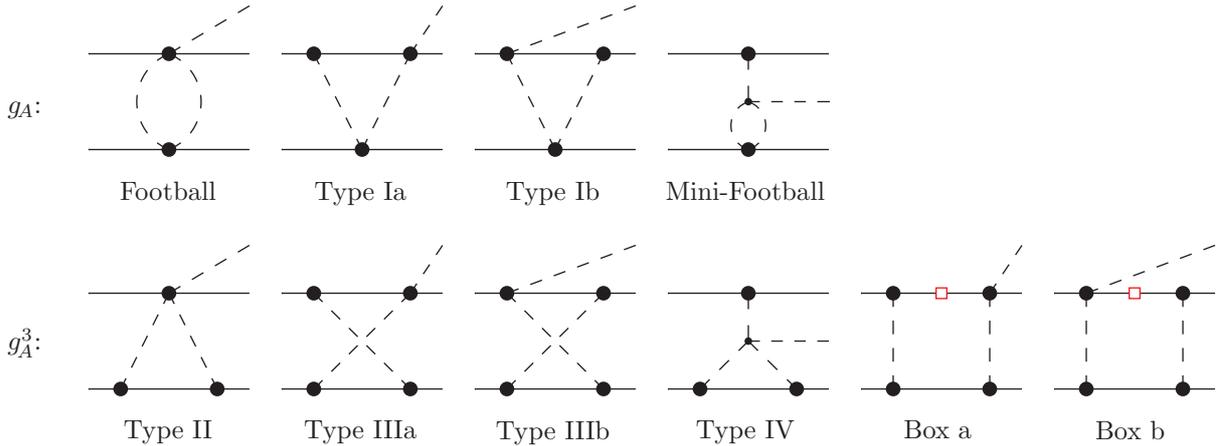}
\caption{\label{fig:allLoops}
One-loop diagrams contributing to s-wave pion production at \NLO{} and \NNLO{}.
Notation is as in Fig.~\ref{fig:allNLO}.
}
\end{figure}
%
This is illustrated by red squares
on the nucleon propagators in the two box diagrams of Fig.~\ref{fig:allLoops}. 
It was shown explicitly in Ref.~\cite{lensky2} that those
induced irreducible contributions
cancel exactly the finite remainder of the NLO loops (II -- IV)
in the $pp\to d\pi^+$ channel.
As a consequence, there are no contributions at NLO  for both
$\pi^0$ and $\pi^+$ productions, see also our results in the two 
first rows of Tables~\ref{tab:ga3} and \ref{tab:ga1}.


In this paper we extend the analysis of the previous studies and
evaluate  the contribution
from pion loops at \NNLO{}.
Once the complete calculation at \NNLO{} is performed, the calculated theoretical
uncertainty based on our power counting
is going to be reduced to   $\sim (\mpi/\mN)^{3/2} < 10\%$ for the amplitudes.
At \NNLO{}, one gets contributions from  two sets of  loop diagrams
which differ in the power of $g_A$.
The diagrams proportional to $g_A^3$ are the subleading contributions
to the NLO diagrams of Fig.~\ref{fig:allNLO} we already discussed.
In addition, there is a set of pion loop diagrams proportional to 
$g_A$, see Fig.~\ref{fig:allLoops}.
A naive MCS estimates indicate that the diagrams proportional to $g_A$ could
play a role already at NLO.
However, a more careful analysis reveals that
the contributions of each of these $g_A$ diagrams at NLO is zero, see
Appendix~\ref{sec:foot} for a detailed discussion.
In subsequent sections it will be shown that 
partial cancellations 
take place among  $g_A$ and $g_A^3$ diagrams at \NNLO{}.
Unlike the cancellation among the $g_A^3$ diagrams at NLO,  
the cancellations  at \NNLO{} are not complete so that 
there is a non-zero transition amplitude from the $g_A$ and $g_A^3$ diagrams.

\begin{figure}[t]
\includegraphics{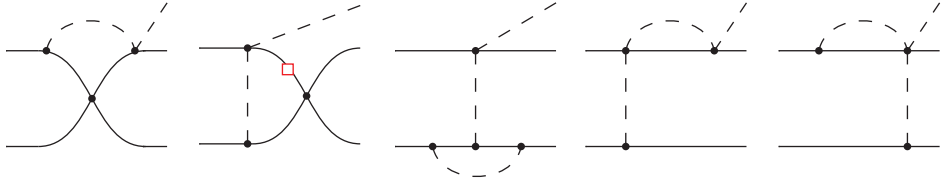}
\caption{\label{fig:highloops}
Exemplary irreducible diagrams  that contribute at higher than \NNLO{}. 
Notation is as in Fig.~\ref{fig:allNLO}.}
\end{figure}

We already discussed pion loop diagrams with the $\pi N\to\pi N$ vertex stemming from
 the leading Weinberg--Tomozawa term, ${\cal L}^{(1)}_{\pi\!N}$,
and its  recoil correction, ${\cal L}^{(2)}_{\pi\!N}$. 
In addition, there are two
kinds of loop diagrams in Fig.~\ref{fig:allLoops}  which involve the
$c_i$-vertices from  ${\cal L}^{(2)}_{\pi\!N}$: those
where the $c_i$ terms appear at the
vertex, where the outgoing on-shell pion is emitted, 
and those where they provide an intermediate interaction. 
The former
kind appears to be suppressed for s-wave pion production due to the pion kinematics
near threshold.
The Lagrangian term containing $c_4$ can not contribute at an outgoing s-wave pion vertex since it
is proportional to the gradient of the pion fields, cf.~Eq.~(\ref{eq:la1}).
The contribution  of the $c_3$ Lagrangian term via 
this type of vertex is only non-zero
if the term proportional to the time derivative of the pion field
is considered in the Lagrangian, cf.~Eq.~(\ref{eq:la1}).
This $c_3$ term, however, results in a  loop amplitude
which is suppressed by $\mpi/\mN$ compared to the leading loop at NLO, and thus
it  is  of higher order (\NNNLO{}).
In addition, the contributions proportional to  LECs $c_1$ and $c_2$ are quadratic with $\mpi$
and therefore strongly suppressed.\footnote{
Naively, the term  proportional $c_2$ scales as $m_\pi (l_0+$recoils$)$. However,
a similar mechanism as the one explained below Eq. (8)  
forces the vertex to become proportional to $m_\pi^2$.
}
However, the contributions of the vertices proportional to $c_2$,
$c_3$ and $c_4$ are potentially important at \NNLO{} once
embedded in the off-shell  intermediate $\pi N$ vertices 
(on the lower nucleon line of the $g_A$-type diagrams in Fig.~\ref{fig:allLoops}).
In what follows we will discuss the individual contributions of loops  in detail.

As a final remark,  we give in Fig.~\ref{fig:highloops} 
some examples of additional loop topologies which start
to contribute at a higher order  than what  is considered in the present study.
The common feature of these diagrams is the presence
of only one pion propagator inside the loops.
As a consequence, by using   appropriate integration
variables, one can eliminate  the large initial 
three-momentum $\vec{p}$ from the loop integrals, meaning 
the loop momentum will scale with $\mpi$. This explains why these 
loop diagrams in Fig.~\ref{fig:highloops} only
start to contribute  at order  N$^3$LO or higher.

\section{Calculation of  diagrams proportional to $\gA^3$}
\label{sec:calcga3}
\begin{figure}[ht]
\includegraphics{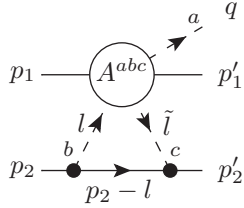}
\caption{\label{fig:ga3struct}
The general structure of $\gA^3$ diagrams. 
}
\end{figure}

The diagrams of the $\gA^3$-group in Fig.~\ref{fig:allLoops}
have a common structure illustrated in
Fig.~\ref{fig:ga3struct}.

The loop diagram in Fig.~\ref{fig:ga3struct}
is integrated over the momentum $l = (l_0,\vec{l})$.
We also use the short-hand notation
\begin{equation*}
\ltild     = l + k_1 - q \; ,
\end{equation*}
with $k_1=p_1-p_1'$.
The pion isospin indices $a$, $b$, and  $c$ are defined as shown in
Fig.~\ref{fig:ga3struct}. The circle  containing the
vertex operator $A^{abc}$ produces an outgoing pion of
isospin index $a$ off nucleon 1.
This operator is
different for each diagram and its explicit form is derived in Appendix~\ref{sec:appga3},
where also the detailed structure of each $g_A^3$-diagram is given.

The invariant amplitude for each relevant diagram proportional to $g_A^3$
can be written as
\begin{equation}
    i M_{\gA^3} = \intdl \Bnotation(l,\ltild ) \ \tau_2^c \tau_2^b \ A_{\gA^3}^{abc} ,
\label{eq:bnotation}
\end{equation}
where $\Bnotation(l,\ltild ) $ is the common operator structure associated with
nucleon 2 in Fig.~\ref{fig:ga3struct}.
The operator   $\Bnotation(l, \ltild) $
involves two pion propagators,
two $\pi NN$-vertices and the nucleon propagator.
The explicit form of $\Bnotation(l,\ltild )$ can be read off from
the diagram in Fig.~\ref{fig:ga3struct}:
\begin{eqnarray}
        \Bnotation(l,\ltild ) &=&
        \frac{i}{l^2 - \mpi^2 + i\nolik} \;
        \frac{i}{\ltild^2 - \mpi^2 + i\nolik} \;
        \frac{i}{p_{20}-l_0-\frac{(\vec p_2 - \vec l)^2}{2 \mN}+i\nolik} \nonumber \\&&\times
        \frac{\gA}{\fpi} \left( -S_2 \cdot \ltild + \frac{S_2 \cdot (\ptwo + \ptwopr - l)
        v \cdot \ltild}{2 \mN} \right)
\nonumber \\ &&\times
        \frac{\gA}{\fpi} \left( S_2 \cdot l - \frac{S_2 \cdot (2 \ptwo - l) v \cdot l}{2 \mN} \right),
\label{eq:defineB}
\end{eqnarray}
where $v^{\mu} = (1,\vec{0})$ is the nucleon four-velocity and $S^{\mu} =(0,{\vec \sigma}/2) $ 
is its spin-vector. 
Note that $\Bnotation(l, \ltild) $ contains no isospin indices as all isospin operators
are included in Eq.~(\ref{eq:bnotation}).
Since the structure of $\Bnotation(l, \ltild) $ is the same for all
considered $g_A^3$ diagrams
we will consider, we
concentrate our discussion on the structure of operator
$A_{\gA^3}^{abc}$ in Eq.~(\ref{eq:bnotation}), see
Appendix~\ref{sec:appga3} for details.
Note that the amplitude, Eq.~(\ref{eq:bnotation}) is not yet properly symmetrized with respect
to the two nucleons.
Below we will first discuss, how the partial cancellation amongst the
various pion loop diagrams emerges on the basis of the decomposition
illustrated in Fig.~\ref{fig:ga3struct}.
In Sec.~\ref{sec:sumTPE} the non-vanishing remainder will be given in
a symmetrized form.

\subsection{Pion s-wave contributions $\propto \gA^3$} 
In Appendix~\ref{sec:appga3} we derive the expressions for each of the six $g_A^3$ diagrams
which contribute to near-threshold
s-wave pion production from two nucleons.
The results of these calculations are summarized in Table~\ref{tab:ga3}
where, for convenience, we have introduced the following short-hand
notation for the  isospin structures:
\begin{equation}
    \taup = (\boldtau_1+\boldtau_2)^a, \qquad
    \taum = (\boldtau_1-\boldtau_2)^a, \qquad
    \taux  = i (\boldtau_1 \times \boldtau_2)^a.
\label{taulabels}
\end{equation}
The left column in Table~\ref{tab:ga3} shows the spin structures that  emerge in these diagrams,
the next six columns  represent the contributions from the individual diagrams to the
given spin structure, whereas the last two columns summarize the net effect of all diagrams
and the MCS order, respectively.
When we add the resulting expressions for the
six diagrams we confirm the finding of Ref.~\cite{lensky2}
that  the sum of the NLO contributions from all diagrams vanishes,
see the first two rows of operators in Table~\ref{tab:ga3}.
Moreover, since the sum of the  operators in the  first two rows of  
Table~\ref{tab:ga3} is an exact zero,
the corresponding spin-momentum structures   $\mystructa$ and  $\mystructb$ 
will  not contribute also
 at N$^2$LO  and all higher orders.
In addition, all nucleon recoil corrections $\propto 1/(2\mN)$ to the individual diagrams
at \NNLO{} also cancel in the sum.
The reason for that cancellation is completely analogous to the cancellation
that happens at NLO, see discussion below Eq.~(\ref{eq:pipivert}).
In fact, only those parts of the $g_A^3$ diagrams 
that cannot be reduced 
to the topology of the  diagram II in Fig.~\ref{fig:allLoops},
give a non-zero contribution to the transition amplitude. 
Thus, only very few \NNLO{} contributions to the pion production amplitude
remain, as seen in Table~\ref{tab:ga3}.
The non-vanishing terms appear from the two cross-box diagrams 
and  diagram  IV.

Since the sum of the $A_{\gA^3}^{abc}$ operators from the different diagrams
starts to contribute at \NNLO{}, we keep only
the leading part of the operator $\Bnotation(l,\ltild )$. 
Adding up the contributions from all six $g_A^3$ diagrams we arrive at the following result:
\begin{eqnarray}
    i M^{\text{\NNLO{}}}_{\gA^3} &=& i\frac{\gA^3}{4 \fpi^5} \intdl
        \frac{ S_2 \cdot \ltild  }{l^2 - \mpi^2 + i\nolik} \;
        \frac{ S_2 \cdot l }{\ltild^2 - \mpi^2 + i\nolik} \;
        \frac{1}{-v\cdot l+i\nolik} \nonumber \\ && \times
        \biggl\{
                (-2\taup+\taux) \frac{2 \vdotq}{-\vdotl + i\nolik}
                   (S_1 \cdot \ltild )
             + (-2\taup-\taux) \frac{2 \vdotq}{-\vdotl + i\nolik}
                  (  S_1 \cdot l) \nonumber  \\ &&
             - 8 \taux (S_1 \cdot k_1 )
                 \frac{ (l+\ltild) \cdot q }{k_1^2-\mpi^2+i\nolik}
        \biggr\},
\label{eq:Mga3first}
\end{eqnarray}
where for the nucleon propagator in Eq.~(\ref{eq:defineB}) we dropped
$p_{20}$ and all recoil terms  of order ${\cal O}(\mpi)$
 compared to the lower-order $l_0 \equiv\vdotl \sim |\vec l|\sim p$ term.
Rearranging the isospin structure we arrive at three independent integrals
to be evaluated for s-wave pion production:
\begin{eqnarray}
     i M^{\text{\NNLO{}}}_{\gA^3} &=& -i \frac{\gA^3}{4 \fpi^5} \nonumber 
            \biggl\{
            4 (\vdotq) \taup S_2^\mu S_2^\nu S_1^\lambda
            \int\limits  \frac{d^4 l}{(2 \pi)^4}  \frac{ \ltild_\mu l_\nu (l+\ltild)_\lambda}
             {(l^2 - \mpi^2 + i\nolik)(\ltild^2 - \mpi^2 + i\nolik)(-\vdotl+ i\nolik)^2} \nonumber \\
            &&-2 (\vdotq) \taux   S_2^\mu S_2^\nu (\sdotk)
            \int\limits \frac{d^4 l}{(2 \pi)^4}  \frac{ \ltild_\mu l_\nu }
                {(l^2 - \mpi^2 + i\nolik)(\ltild^2 - \mpi^2 + i\nolik)(-\vdotl+ i\nolik)^2} \nonumber \\
            &&+ 8 q^\lambda  \taux
                    \frac{S_2^\mu S_2^\nu (\sdotk)}{k_1^2-\mpi^2+i\nolik}
            \int\limits \frac{d^4 l}{(2 \pi)^4}  \frac{ \ltild_\mu l_\nu (l+\ltild)_\lambda}
                               {(l^2 - \mpi^2 + i\nolik)(\ltild^2 - \mpi^2 + i\nolik)(-\vdotl+ i\nolik)}
            \biggr\}
\label{eq:Mga3int}
\end{eqnarray}

\begingroup
\squeezetable
\begin{table}[ht]
\caption{\label{tab:ga3} Interference pattern of \NLO{} and \NNLO{} s-wave contributions
from the individual $\gA^3$ diagrams. The  Table shows the contributions to  the 
vertex $\tau_2^c\tau_2^b \ A_{\gA^3}^{abc}$ defined in Eq.\eqref{eq:bnotation} and
Fig.~\ref{fig:ga3struct}.
 These contributions are given separately 
for the different spin-momentum  structures of the vertex $A_{\gA^3}^{abc}$, 
shown in the leftmost column. The notation for the isospin structures is
defined in Eq.\eqref{taulabels}.
}
\begin{ruledtabular}
\begin{tabular}{lcccccccr}
               &       Type II       &     Type IIIa   &    Type IIIb   &    Type IV       &    Box a      &    Box b      &  Sum         & Order   \\ \hline
 $ \mystructa $&$   -4\zp-4\zm+2\zx $&$  0            $&$   -2\zp-\zx  $&$\pp6\zp+6\zm    $&$  -2\zm-\zx  $&$   0         $&$  0         $& \NLO{}, \NNLO{}  \\
 $ \mystructb $&$ \pp4\zp+4\zm+2\zx $&$ \pp2\zp-\zx   $&$    0         $&$  -6\zp-6\zm    $&$   0         $&$\pp2\zm-\zx  $&$  0         $& \NLO{}, \NNLO{}  \\
 $ \mystructc $&$   -2\zp+2\zm      $&$ \pp2\zp-\zx   $&$    0         $&$   0            $&$   0         $&$  -2\zm+\zx  $&$  0         $& \NNLO{} \\
 $ \mystructd $&$ \pp2\zp-2\zm      $&$  0            $&$   -2\zp-\zx  $&$   0            $&$\pp2\zm+\zx  $&$   0         $&$  0         $& \NNLO{} \\
 $ \mystructe $&$ \pp4\zp+4\zm-2\zx $&$  0            $&$ \pp2\zp+\zx  $&$  -6\zp-6\zm    $&$\pp2\zm+\zx  $&$   0         $&$  0         $& \NNLO{} \\
 $ \mystructf $&$   -4\zp-4\zm-2\zx $&$ -2\zp+\zx     $&$    0         $&$\pp6\zp+6\zm    $&$   0         $&$  -2\zm+\zx  $&$  0         $& \NNLO{} \\
 $ \mystructg $&$    0              $&$  0            $&$   -2\zp-\zx  $&$   0            $&$   0         $&$   0         $&$  -2\zp-\zx $& \NNLO{} \\
 $ \mystructh $&$    0              $&$ -2\zp+\zx     $&$    0         $&$   0            $&$   0         $&$   0         $&$  -2\zp+\zx $& \NNLO{} \\
 $ \mystructi $&$    0              $&$  0            $&$    0         $&$  -8\zx         $&$   0         $&$   0         $&$  -8\zx     $& \NNLO{} \\
\end{tabular}
\end{ruledtabular}
\end{table}
\endgroup

Employing dimensional regularization and an integration method
outlined in
Appendix~\ref{sec:appintegr},
Eq.~(\ref{eq:Mga3int}) can be brought into the more transparent form:
\begin{eqnarray}
    i M^{\text{\NNLO{}}}_{\gA^3} &=&  \frac{\gA^3 (v \cdot q)}{\fpi^5} \bigg\{
    \taup  i \varepsilon^{\mu\nu\alpha\beta}
       k_{1\mu} S_{1\nu} v_\alpha S_{2\beta}
    \left[ - J(k_1^2)\right]
\nonumber \\
    &&
+ \taux (S_1 \cdot k_1)
    \left[ -\frac{19}{24}  J(k_1^2) + \frac{5}{9}  \frac{1}{(4 \pi)^2} \right]
    \bigg\},
\label{eq:Mga3result}
\end{eqnarray}
where we have only kept the lowest  order parts of the
integrals which give contributions to the amplitude at \NNLO{}.
The pion loop diagrams generate ultraviolet divergent terms,
which are contained in the following integral:
\begin{equation}
J (k_1^2) =-i
\int\limits  \frac{d^4 l}{(2 \pi)^4}
\frac{1}{l^2-\mpi^2+i0} \frac{1}{(l+k_1)^2-\mpi^2+i0}
\label{eq:Ipi}
\end{equation}
The divergences are to be absorbed
by the LECs accompanying the five-point ($4N\pi$) vertices 
as we will discuss in Sec.~\ref{sec:reg}.

%
\section{Calculation of diagrams proportional to  $\gA$ }
\label{sec:calcga1}
\begin{figure}[ht]
\includegraphics{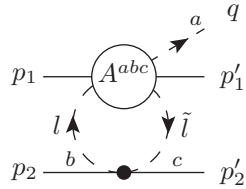}
\caption{\label{fig:ga1struct}The general structure of $\gA$ diagrams.}
\end{figure}

We evaluate the  $\gA$ diagrams following a similar strategy we used
when we evaluated the $\gA^3$ diagrams.
The invariant amplitude for each diagram proportional to $g_A$
can be written as
\begin{equation}
    i M_{\gA} = \intdl \Dnotation(l,\ltild ) \ \varepsilon^{bcd}\tau_2^d  \ A_{\gA}^{abc} ,
\label{eq:MgaGeneral}
\end{equation}
where $\Dnotation(l,\ltild )$ is a common operator structure which is associated with
nucleon 2 in Fig.~\ref{fig:ga1struct}.

This structure involves the WT vertex at the second nucleon and the two pion propagators:
\begin{equation}
        \Dnotation(l,\ltild)  =
        \frac{i}{l^2 - \mpi^2 + i\nolik} \;
        \frac{i}{\ltild^2 - \mpi^2 + i\nolik} \;
        \frac{      v \cdot (l + \ltild)}{4\fpi^2}.
\label{eq:defineD}
\end{equation}
Note that we have  only written  the leading WT-vertex contribution
in  $\Dnotation(l,\ltild)$, Eq.~(\ref{eq:defineD}), 
since the sum of the $ A_{g_A}^{abc}$ operators
starts to contribute at \NNLO{} only, as can be seen in Table~\ref{tab:ga1}. 
This is 
in full analogy to  the sum of the $ A_{g_A^3}^{abc}$ operators, 
which also only start to contribute at \NNLO{}, where, as discussed just before
Eq.~(\ref{eq:Mga3first}), the recoil ($1/m_N$) 
corrections in $B_2(l,\ltild)$, Eq.~(\ref{eq:defineB}), only contribute at higher order. 
In other words, 
the corrections to  $\Dnotation(l,\ltild)$, 
that is
the recoil correction to the leading WT interaction term
and the correction stemming from the $c_4$-vertex,
contribute at a higher order than what is considered in this work.
Notice further that the $c_2$ and $c_3$-vertices   in Eq.~(\ref{eq:la1}) are 
isoscalars and, therefore, do not contribute to  the function
$\Dnotation(l,\ltild)$. The  contributions of these LECs  will be discussed in the next section.

\subsection{Pion s-wave contributions $\propto \gA$}
The operator expressions for  each individual diagram of the $g_A$-type contributing
to s-wave  pion  production can be found in
Appendix~\ref{sec:appga1}. In a complete analogy with the $g_A^3$ diagrams,
we summarize in Table~\ref{tab:ga1} the contributions of  the individual
diagrams and their net effect for different spin structures.
In distinction to the $g_A^3$-graphs  
the diagrams of this topology do not appear, contrary to naive MCS expectations, at NLO,
see Appendix~\ref{sec:foot} for a more detailed discussion.
Similarly to $g_A^3$-type contributions, only a few of the \NNLO{}
terms do not cancel in the sum.
Again, only those parts of the diagrams Ia, Ib and  mini-football 
that cannot be reduced 
to the topology of the football diagram in Fig.~\ref{fig:allLoops},
give a non-zero contribution to the transition amplitude. 
The results are shown in Table~\ref{tab:ga1}, and the sum of these 
$g_A$ contributions gives the following transition amplitude:
\begin{eqnarray}
        i M^{\text{\NNLO{}}}_{\gA} &=& i
               \frac{\gA}{8 \fpi^3} \intdl \Dnotation(l,\ltild)\
            \biggl\{
              (2\taup -2\taum - 2\taux) \frac{2\vdotq}{- \vdotl +i \nolik} \sdotltild \nonumber \\
             && + (2\taup - 2\taum + 2\taux) \frac{2\vdotq}{- \vdotl +i \nolik} \sdotl+
              8 \taux\; \sdotk \frac{ q \cdot (l + \ltild)}{k_1^2-\mpi^2+i\nolik}
            \biggr\}.
\label{eq:MgaD}
\end{eqnarray}

\begingroup
\squeezetable
\begin{table}[bt]
\caption{\label{tab:ga1} Interference pattern of s-wave contributions
from the individual $\gA$ diagrams. The  Table shows the contributions to  the 
vertex $\varepsilon^{bcd}\tau_2^d A_{\gA}^{abc}$ defined in Fig.~\ref{fig:ga1struct} and
Eq.\eqref{eq:MgaGeneral}. 
 The contributions are given separately 
for different spin-momentum  structures of the vertex $A_{\gA}^{abc}$, shown
in the leftmost column.
 The notation for the isospin structures is
defined in Eq.\eqref{taulabels}.
}
\begin{ruledtabular}
\begin{tabular}{lcccccr}
                             &  Football     &  Type Ia          &     Type Ib      &  MiniFB &   Sum             & Order   \\ \hline
$ \mystructa $&$  -2\zx       $&$  -\zp+\zm+\zx    $&$\pp\zp-\zm+\zx  $&$ 0     $&$  0              $& NLO\footnotemark[1], \NNLO{}\\
$ \mystructb $&$  -2\zx       $&$  -\zp+\zm+\zx    $&$\pp\zp-\zm+\zx  $&$ 0     $&$  0              $& NLO\footnotemark[1], \NNLO{}\\
$ \mystructc $&$\pp2\zp-2\zm  $&$  -\zp+\zm+\zx    $&$  -\zp+\zm-\zx  $&$ 0     $&$  0              $& \NNLO{} \\
$ \mystructd $&$  -2\zp+2\zm  $&$\pp\zp-\zm-\zx    $&$\pp\zp-\zm+\zx  $&$ 0     $&$  0              $& \NNLO{} \\
$ \mystructe $&$\pp2\zx       $&$\pp\zp-\zm-\zx    $&$  -\zp+\zm-\zx  $&$ 0     $&$  0              $& \NNLO{} \\
$ \mystructf $&$\pp2\zx       $&$\pp\zp-\zm-\zx    $&$  -\zp+\zm-\zx  $&$ 0     $&$  0              $& \NNLO{} \\
$ \mystructg $&$\pp0          $&$   0              $&$ 2\zp-2\zm+2\zx $&$ 0     $&$  2\zp-2\zm+2\zx $& \NNLO{} \\
$ \mystructh $&$\pp0          $&$  2\zp-2\zm-2\zx  $&$   0            $&$ 0     $&$  2\zp-2\zm-2\zx $& \NNLO{} \\
$ \mystructi $&$\pp0          $&$   0              $&$   0            $&$ 8\zx  $&$  8\zx           $& \NNLO{} \\
\end{tabular}
\end{ruledtabular}
\footnotetext[1]{Notice that, in addition to the cancellation
shown in the Table, in the case of NLO even the individual contributions
to the corresponding spin-momentum structures turn out to vanish,
see Appendix~\ref{sec:foot} for details.}
\end{table}
\endgroup

We now turn to the contribution emerging from the diagrams of Fig.~\ref{fig:allLoops} with the
$c_2$ and $c_3$-vertices in the  off-shell pion kinematics at nucleon 2. We
obtain the following expression for the amplitude:
\begin{eqnarray}\nonumber
        i M^{\text{\NNLO{}}}_{\gA, c_i} &=& -i \frac{\gA}{2 \fpi^5} (\taup + \taum) (S\cdot  k_1)\\ &\times &
        \intdl \frac{ c_3(l \cdot \ltild)+(c_2-g_A^2/8m_N) (v \cdot l)(v \cdot \ltild) }{(l^2 - \mpi^2 + i\nolik)
(\ltild^2 - \mpi^2 + i\nolik)} \;
\; \biggl\{ 2  +\frac1{2}+ \frac1{2}- 3  \biggr\}=0 \; ,
\label{eq:c3}
\end{eqnarray}
where the numbers in the curly bracket correspond to the individual contributions of 
the $g_A$-diagrams, as they 
appear in Fig.~\ref{fig:allLoops},  in order. 
Again, while the individual diagrams do contribute at \NNLO{}, their
sum turns out to yield a vanishing result.
We, therefore, conclude that there are no loop amplitudes
$\propto c_i$  to the order we are working.

Upon performing some simplifications, the total result for the
$g_A$-contribution to the transition amplitude in
Eq.~(\ref{eq:MgaD}) can be brought into the form
\begin{eqnarray}
        i M^{\text{\NNLO{}}}_{\gA} &=& -i\frac{\gA}{8 \fpi^5}
        \biggl\{
         (\taup - \taum) (\vdotq)
             \intdl  \frac{v \cdot (l+\ltild) S_1 \cdot (l+\ltild)}
         {(l^2 - \mpi^2 + i\nolik)(\ltild^2 - \mpi^2 + i\nolik)(-\vdotl+ i\nolik)} \nonumber \\
         && - \taux (\vdotq) (\sdotk)
             \intdl  \frac{v \cdot (l+\ltild)}
         {(l^2 - \mpi^2 + i\nolik)(\ltild^2 - \mpi^2 + i\nolik)(-\vdotl+ i\nolik)} \nonumber \\
         && + 2 \taux (\sdotk) \frac{1}{k_1^2-\mpi^2+i\nolik}
             \intdl  \frac{v \cdot (l+\ltild)  q \cdot (l+\ltild)}
         {(l^2 - \mpi^2 + i\nolik)(\ltild^2 - \mpi^2 + i\nolik)}
        \biggr\} \,.
\label{eq:Mga1int}
\end{eqnarray}
The first term in Eq.~(\ref{eq:Mga1int}) does not contribute at \NNLO{},
since at this order the term $v \cdot (l + \ltild)\approx 2v \cdot l$
in the numerator cancels
with the nucleon propagator $-v \cdot l + i \nolik$.
The resulting integral vanishes due to the
symmetry of the integrand.
Specifically, the integral is to be invariant under the shift of variables
($l\to -\tilde l, \tilde l \to -l$). Indeed,
the denominator of  this integrand is invariant under this transformation
 whereas the numerator changes its sign. Therefore, the first term in
Eq.~(\ref{eq:Mga1int}) is equal to zero.
Finally,  keeping only the lowest-order terms as appropriate at \NNLO{} and
using the expressions for the loop integrals outlined in Appendix~\ref{sec:appintegr},
we arrive at the final result:
\begin{equation}
    i M^{\text{\NNLO{}}}_{\gA} =
    \frac{\gA}{\fpi^5} \taux (v \cdot q) (S_1 \cdot k_1)
     \left[ \frac16 J(k_1^2) - \frac{1}{18} \frac{1}{(4 \pi)^2} \right]\,,
\label{eq:Mga1result}
\end{equation}
where the UV-divergent integral $J(k_1^2)$  is defined in Eq.~(\ref{eq:Ipi}).

\section{Summary of the two-pion exchange diagrams }
\label{sec:sumTPE}
Until now we
have evaluated the expressions for the production operator assuming
that the pion is produced from nucleon 1. We now add the contribution
emerging from interchanging the nucleon labels.  We use the fact that in the
center-of-mass system $\vec{p}_1 =
-\vec{p}_2 = \vec{p}$ and $k_1=-k_2+q$ and employ the approximate relation $k_1^2 \simeq k_2^2$
with higher-order terms being ignored.  Throughout, we also ignore
operators leading to pion p-wave production.  We then obtain  from
Eqs.~(\ref{eq:Mga1result}) and (\ref{eq:Mga3result})  the following
complete (i.e. symmetrized with respect to the nucleon labels)
expressions:
\begin{equation}
i M^{\text{\NNLO{}}}_{\gA} = \frac{\gA \ (v \cdot q)}{\fpi^5}
\taux  (S_1 + S_2)\cdot k_1
\left[ \frac16 J(k_1^2) - \frac{1}{18} \frac{1}{(4 \pi)^2} \right] \; ,
\label{eq:Mga1final}
\end{equation}
\begin{eqnarray}
    i M^{\text{\NNLO{}}}_{\gA^3} &=& \frac{\gA^3 \ (v \cdot q)}{\fpi^5} \bigg\{
    \taup i \varepsilon^{\alpha\mu\nu\beta} v_\alpha k_{1\mu} S_{1\nu} S_{2\beta}
    \left[-2 J(k_1^2)\right] \nonumber \\
    && + \taux  (S_1 + S_2) \cdot k_1
    \left[ -\frac{19}{24}  J(k_1^2) + \frac{5}{9} \frac{1}{(4 \pi)^2} \right]
    \bigg\}.
\label{eq:Mga3final}
\end{eqnarray}
Employing dimensional regularization, $d=4-\varepsilon$), the integral $J(k_1)$
entering the above expressions can be written in the form
\begin{eqnarray}
J (k_1^2) &=&
\frac{\mu^{\varepsilon}}{i} \int \frac{d^{(4-\varepsilon)} l}{(2 \pi)^{(4-\varepsilon)}}
\frac{1}{[l^2-\mpi^2+i0][(l+k_1)^2-\mpi^2+i0]} \nonumber \\
& =& -2 L -
\frac{1}{(4 \pi)^2} \left [ \log \left( \frac{\mpi^2}{\mu^2} \right)
- 1 + 2 F_1 \left(\frac{k_1^2}{\mpi^2}\right) \right],
\label{eq:Ipi_final}
\end{eqnarray}
where the function $F_1 (x)$ is defined via
\begin{equation}
\newcommand{\myvar}{x}
F_1 (\myvar) = \frac{\sqrt{4-\myvar-i0}}{\sqrt{\myvar}}
\arctan \left( \frac{\sqrt{\myvar}}{\sqrt{4-\myvar-i0}} \right)\,. 
\label{eq:F1}
\end{equation}
and  the UV divergency appears as a simple pole in the function $L$:
\begin{equation}
L= \frac{1}{(4\pi)^2}\left[- \frac{1}{\varepsilon}+\frac{1}{2}
\left(\gamma_E-1-{ \log}(4\pi)\right)\right].
\end{equation}
Note that both $M^{\text{\NNLO{}}}_{\gA}$ and $M^{\text{\NNLO{}}}_{\gA^3}$
are proportional to the outgoing pion energy $v\cdot q \simeq \mpi$, i.e.
both operator amplitudes vanish at threshold in the chiral limit.

\section{Regularization procedure}
\label{sec:reg}
In MCS the loop diagrams which contribute to the renormalization of
e.g. the nucleon mass $\mN$ and the
axial coupling constant $g_A$ do not involve large-momentum
components.
Consequently, these diagrams contribute in the MCS at order \NNNNLO{}
which is beyond the scope of the present work.
For example, consider a LO rescattering diagram which in our naive
counting is of order $\sqrt{\mpi / \mN}$. 
Including a pion loop in any of these diagram  
will require a renormalization any of the
vertices in these LO diagrams, cf. e.g. the last three diagrams in 
Fig.~\ref{fig:highloops}. 
This pion loop will increase the MCS order
by  factor $(\mpi / \mN)^2$ as shown in, e.g.~Ref.~\cite{hanhart04}, Table 11.
At \NNLO{}, we only have to consider the loop diagrams which are evaluated in this paper.
The UV divergences appearing in the corresponding integrals are to be
absorbed into  LECs accompanying the $4N\pi$ amplitudes 
${\cal A}_\text{CT}$ and ${\cal B}_\text{CT}$ introduced in Sec.~\ref{sec:ampl}.
The contributions of the loops to the amplitudes
${\cal A}$ and ${\cal B}$, see Eq.~\eqref{eq:Mthr},
can be separated into  singular and  finite parts
\begin{eqnarray}
{\cal A}&=& \frac{\mpi}{(4\pi \fpi)^2\fpi^3}(\tilde {\cal A}_\text{singular}
+\tilde{\cal A}_\text{finite} ), \nonumber \\
{\cal B}&=&\frac{\mpi}{(4\pi \fpi)^2\fpi^3} (\tilde {\cal B}_\text{singular}
+\tilde {\cal B}_\text{finite} ), 
\label{eq:loops1}
\end{eqnarray}
where
\begin{eqnarray}
\tilde {\cal A}_\text{singular}= g_A^3 (4\pi)^2 L,   \quad \quad
\tilde {\cal B}_\text{singular}=
-\frac{g_A}{6}\left( \frac{ 19}{4} g_A^2 -1 \right) (4\pi)^2 L \; .
\end{eqnarray}
Here we have used that at threshold $\vec{k}_1=\vec{p}$ and  $v\cdot q = \mpi$.
Notice that the above decomposition into singular and finite pieces is, clearly, scheme dependent.
Analogously, the amplitudes given by the  $4N\pi$ Lagrangian contact  terms, which are 
given in, e.g., Ref.~\cite{cohen}, are  written as:
\be
{\cal A}_\text{CT}=\frac{\mpi}{(4\pi \fpi)^2\fpi^3}(\tilde{\cal A}^r_\text{CT}(\mu)
+(4\pi)^2 \beta_{\cal A}L), \quad
{\cal B}_\text{CT}=\frac{\mpi}{(4\pi \fpi)^2\fpi^3}(\tilde{\cal B}^r_\text{CT}(\mu)
+(4\pi)^2 \beta_{\cal B}L) \; .
\label{eq:CT}
\ee
The singular  parts of the amplitudes in Eq.~(\ref{eq:CT})  cancel
the singularities of the amplitudes in Eq.~(\ref{eq:loops1}), 
emerging from the loops.  
The resulting finite expressions for the scattering amplitudes 
are given in terms of the 
renormalized LECs of Ref.~\cite{cohen}.
\begin{eqnarray}
{\cal A}^r_\text{CT}&=& \frac{\mpi}{(4\pi \fpi)^2\fpi^3} \tilde{\cal A}^r_\text{CT}=
 -(d_1'+2e_1-2e_2)\frac{\mpi}{4m_N\fpi} 
\nonumber \\  
{\cal B}^r_\text{CT}&=& \frac{\mpi}{(4\pi \fpi)^2\fpi^3} \tilde{\cal B}^r_\text{CT}=
 -(d_1'+2e_1)\frac{\mpi}{4m_N\fpi} 
\label{eq:CT-resonance}
\end{eqnarray} 
The magnitudes of the amplitudes ${\cal A}^r_\text{CT}$ and  ${\cal B}^r_\text{CT}$
can be estimated 
using the values of the LECs determined in  Refs.~\cite{cohen,rocha,vmr96}
where the short-ranged production mechanisms
were assumed to originate from z-diagrams with $\sigma$ and $\omega$ exchanges
(see explicit expressions for these exchanges in Refs.~\cite{cohen,ksmk09}). 
Given the estimates in Ref.~\cite{vmr96}, we find $d_1'+2e_1-2e_2 \simeq - 7.5/f^2_\pi
\mN$ and $d_1'+2e_1 \simeq - 3.5/f^2_\pi \mN$, and 
using   $m_N \simeq 4\pi f_\pi $, we obtain  
${\cal A}^r_\text{CT}\simeq 2\, {\mpi}/{(m_N^2\fpi^3)}$ and 
${\cal B}^r_\text{CT}\simeq 1 \, {\mpi}/{(m_N^2\fpi^3)}$, 
which results in $\tilde{\cal A}^r_\text{CT}\simeq 2$ and $\tilde{\cal B}^r_\text{CT}\simeq 1$.

We take these numbers to set the scale for typical \NNLO{} contributions.
Therefore, these estimates allow us to infer the importance of the pion--nucleon loop contributions
to the $NN\to NN\pi$ reactions at threshold.
In particular, we can compare
this estimate with the finite parts of the loops given by Eqs.~(\ref{eq:Mga1final}) and
(\ref{eq:Mga3final}) 
(where $v\cdot p  \sim m_\pi \ll |\vec{p} |$).
\begin{eqnarray}
\label{ABfin}
\tilde {\cal A}_\text{finite}&=& -\frac{g_A^3}{2}
\left[ 1 - \log \left( \frac{\mpi^2}{\mu^2} \right)
-  2 F_1 \left(\frac{-{\vec p\,}^2}{\mpi^2}\right) \right],
\nonumber \\
 \tilde {\cal B}_\text{finite}&=& -\frac{g_A}{6}\left[ -\frac{1}{2}\left(\frac{ 19}{4} g_A^2 -1\right)
\left( 1 - \log \left( \frac{\mpi^2}{\mu^2} \right)
-  2 F_1 \left(\frac{-{\vec p\,}^2}{\mpi^2}\right) \right)+ \frac{5}{3}g_A^2-\frac16
 \right] \ ,
\end{eqnarray}
Choosing $\mu=4\pi \fpi$ with $\fpi=92.4$ MeV  and $g_A=1.32$,  we find 
$\tilde {\cal A}_\text{finite}=-2.9$ and $\tilde {\cal B}_\text{finite}=1.4$.
We, therefore, conclude that contributions of the finite parts of the
loops are comparable in size with
$\tilde{\cal A}^r_\text{CT}$ and $\tilde{\cal B}^r_\text{CT}$. This confirms our power counting
and shows that pion loops contributions, not considered in previous
analyses, are indeed significant.
One should, however, keep in mind that this result was obtained for
the particular regularization scheme as explained above.
In general, the finite parts of the loops 
$ \tilde {\cal A}_\text{finite}$ and  $ \tilde {\cal B}_\text{finite}$
can be further decomposed into the {\it short}- and {\it long}-range parts.
The former one is just a (renormalization scheme dependent) constant 
to which  all terms in Eq.~\eqref{ABfin} but $F_1$ contributes.
On the other hand, the long-range part of the loops is scheme-independent. 
By expanding the function $F_1(-{\vec p\,}^2/m_\pi^2)$, Eq.~(\ref{eq:F1}), 
which is the only long-range piece in 
\eqref{ABfin}, in the kinematical
regime relevant for pion production, i.e.~$({\vec p\,}^2/m_\pi^2) \gg 1$, 
up to the terms at \NNLO{} one obtains
\be
\tilde {\cal A}^\text{long}_\text{finite}&=&-\frac{g_A^3}{2} \log\left(\frac{\mpi^2}{{\vec p\,}^2}\right) + 
{\cal O}\left(\frac{\mpi^2}{{\vec p\,}^2}\right),
\nonumber \\
 \tilde {\cal B}^\text{long}_\text{finite}&=& \frac{g_A}{12}\left(\frac{ 19}{4} g_A^2 -1\right)
\log\left(\frac{\mpi^2}{{\vec p\,}^2}\right)+ {\cal O}\left(\frac{\mpi^2}{{\vec p\,}^2}\right).
\ee
Numerical evaluation of these terms gives 
$ \tilde {\cal A}^\text{long}_\text{finite}=2.2$ and 
$ \tilde {\cal B}^\text{long}_\text{finite}=-1.5$.
The scheme-independent long range part of \NNLO{} pion loops 
 appears to be as large as the  resulting short-range amplitudes, 
$\tilde{\cal A}^r_\text{CT}$ and $\tilde{\cal B}^r_\text{CT}$, 
which are given by the meson-exchange mechanism, 
proposed in 
Refs.~\cite{Lee,HGM,Hpipl,jounicomment} 
to resolve the discrepancy between phenomenological 
calculations and experimental data.
Hence, the importance of the \NNLO{} pion loop effects, 
not included in the previous studies, 
raises serious doubts on the physics interpretation 
behind the phenomenologically  successful models  of
Refs.~\cite{Lee,HGM,Hpipl,jounicomment}. 

In a subsequent work we will present results for \NNLO{} loops 
including the Delta resonance as well as the convolution with
proper nuclear wave functions. At that point a fit to the pion
production data is possible and we can extract the strength of the
counter terms from data.

\section{Summary and discussion}
\label{sec:outlook}

Chiral perturbation theory  has been successfully applied in the past
decades to describe low-energy dynamics of pions and nucleons.
Application of this theoretical framework to pion production in
nucleon-nucleon collisions is considerably more challenging due to
the large three-momentum transfer involved in this reaction. The
slower convergence of the chiral expansion for this reaction, i.e.~the
expansion in the parameter $\chi \sim \sqrt{m_\pi / m_N}$  defined
in Eq.~(\ref{expansionpapar}), provides a strong motivation for
extending the calculations to higher orders. 
In this work we used 
the power counting scheme which properly accounts for the additional scale
associated with the large momentum transfer, namely the momentum
counting scheme (MCS),  to classify various contributions to the
$NN\to NN\pi$ transition amplitudes according to their importance. We
also evaluated all loop diagrams with pions and nucleons as the only
explicit degrees of freedom up to and including \NNLO{}.
The considered loop diagrams can be divided into two groups according
to the power of the nucleon axial-vector coupling constant:  the ones
linear with $g_A$  and the ones proportional
to $g_A^3$, see Fig.~\ref{fig:allLoops}. We confirm the earlier
findings that there are no  NLO loop contributions to the
threshold $NN\to NN\pi$ reaction amplitudes.
Our results, which are
partially summarized in Tables~\ref{tab:ga3} and \ref{tab:ga1} and in 
Secs.~\ref{sec:sumTPE} and \ref{sec:reg},
demonstrate that the MCS combined with the requirements of chiral symmetry
(breaking) pattern of QCD lead to a high degree of 
cancellation among various \NNLO{} contributions.
In particular, all $1/m_N$-corrections of the various diagrams cancel at \NNLO{}.
We also show that the LECs $c_i$, $i=1\ldots  4$,
of ${\cal L}^{(2)}_{\pi\!N}$ do not contribute to the pion loops at this order.

From Table~\ref{tab:ga3}
we see that only the cross-box diagram (diagram III)
and the four-pion interaction diagram (diagram  IV)
contribute to the pion s-wave transition amplitude
$M_{g_A^3}^\text{\NNLO{}}$ given in Eq.~(\ref{eq:Mga3final}).
The two cross-box diagrams
contribute to both amplitudes, the isoscalar one ${\cal A}$  and
the isovector one ${\cal B}$,
whereas diagram IV only contributes to ${\cal B}$.
Analogously, from Table~\ref{tab:ga1} one can deduce that the non-vanishing contributions
to the amplitude $M_{g_A}^\text{\NNLO{}}$,
Eq.~\eqref{eq:Mga1final}, 
originate from the double $\pi N$ scattering diagrams of type Ia and Ib and
from the mini-football diagram. These diagrams  
however contribute only to the isovector amplitude ${\cal B}$, 
as seen in Eq.~(\ref{eq:Mga1final}).
Thus the only contribution from pion loops to the isoscalar amplitude, ${\cal A}$,  
originates from the cross-box diagrams.

The pattern of cancellations discussed above has important phenomenological
implications. In fact, none of the previous phenomenological
investigations take into
account either the cross box diagrams (type III) or the double scattering
contributions (type I) which, as we find, contribute significantly to
 the production amplitude. In particular, the
regularization-scheme independent  long-range contribution of the pion
loops to  ${\cal A}$ turns out to be comparable in size with the
short-range amplitudes emerging in phenomenological models of 
Refs.~\cite{Lee,HGM,Hpipl,jounicomment}
from heavy-meson z-diagrams which, in these studies, are advocated as
the necessary mechanism to describe experimental data. 
Thus, our
findings raise doubts on the role  of the
short-range physics in pion production as suggested in these
phenomenological studies. We, however, refrain from making a more
definite conclusions until the complete
\NNLO{} operator convoluted with the nucleon wave functions 
is confronted with experimental data~\cite{future}.

\begin{figure}[tb!]
\includegraphics{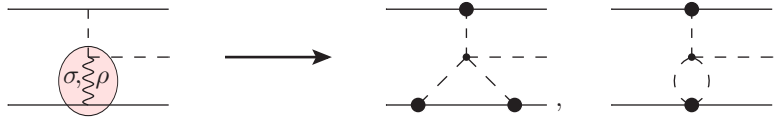}
\caption{\label{phenom}
Diagram on the left-hand-side represents the phenomenological rescattering mechanism 
via the off shell $\pi N$-amplitude.
Diagrams on the right-hand-side are the only loop graphs at N$^2$LO that  can be
interpreted as an analog of this phenomenological mechanism. For
notation see Fig.~\ref{fig:allNLO}.}
\end{figure}

Meanwhile,  within various meson-exchange approaches~\cite{eulogio,unsers,mosel},
the pion production is largely driven by tree-level pion rescattering off a nucleon
with the  $\pi N\to \pi N$ amplitude being far off shell, see Fig.~\ref{phenom}. 
The physics
associated with $\pi N$ scattering  near threshold is normally
parameterized in phenomenological calculations
in terms of the $\sigma$- and $\rho$-meson-exchange contributions. 
The
scalar-isoscalar ($\sigma$-type) $\pi N$ interaction
is relevant for the isoscalar production amplitude  ${\cal A}$ while the isovector
($\rho$-type) $\pi N$ interaction contributes to the strength of  ${\cal B}$. 
The isoscalar  $\pi N$ scattering
amplitude essentially vanishes on-shell, see Refs.~\cite{ourpid} for 
the most recent evaluation of
the isoscalar $\pi N$ scattering length. 
Therefore, the mechanism of Refs.~\cite{eulogio,unsers,mosel}
relies on the  significance  of the off-shell properties of the $\pi N$ scattering
amplitude.
Our EFT consideration puts this mechanism into question.
Pion rescattering via the  phenomenological  pion-nucleon transition amplitude
can in chiral EFT 
be mapped  onto  pion rescattering (at tree level)
via the low-energy constants $c_i$  
plus some contributions from pion loops.
The tree-level piece $\propto c_i$ is, even in the off-shell (pion
production) kinematics by far too small to explain the data for the
neutral pion production \cite{cohen,park}.
As far as the loop contributions are concerned, only diagrams IV and
mini-football may be regarded as an analog
of the corresponding phenomenological mechanism, as illustrated in  Fig.~\ref{phenom}. 
However, after the cancellations, the only contribution that survives
from diagram IV has an isovector structure
as shown in Table~\ref{tab:ga3}.
Furthermore, the mini-football diagram gives an isovector contribution, 
see Table~\ref{tab:ga1}.
Therefore, none of the pion loops can be mapped 
into the
particular phenomenological mechanism in the isoscalar case.
Thus, another lesson we learn from our work about the phenomenology of the neutral pion
production near threshold is that the rescattering contribution with
the isoscalar $\pi N$ amplitude modeled phenomenologically 
by a $\sigma$ exchange should be very small.
On the other hand, the rescattering mechanism  with
the  isovector $\pi N$ amplitude,  the Weinberg--Tomozawa term related to the 
$\rho$-meson exchange
via Kawarabayashi--Suzuki--Riazuddin--Fayyazuddin relation \cite{KS,RF},
is potentially capable of resolving the discrepancy with the experimental
data for charged pion production \cite{lensky2}.

\section*{Acknowledgments}

This work was
supported by funds provided by the Helmholtz Association (VH-VI-231),  
the EU HadronPhysics3 project ``Study
of strongly interacting matter'', the European Research Council
(ERC-2010-StG 259218 NuclearEFT),  DFG-RFBR grant (436 RUS 113/991/0-1),
and the National Science Foundation (US) grants PHY-0758114 and PHY-1068305. 

\appendix


\section{The evaluation of the individual $g_A^3$-diagrams}
\label{sec:appga3}
In this appendix we derive the NLO and \NNLO{}  expressions  for individual
two-pion exchange diagrams shown in Fig.~\ref{fig:allLoops} under restriction that the outgoing pion is
produced in s-wave. The kinematics is defined in Fig.~\ref{fig:ga3struct}.

\subsection{Diagram II}

Diagram  II shown in Fig.~\ref{fig:allLoops} is straightforward to evaluate.
The operator $A^{abc}_{g_A^3}$ of Eq.~(\ref{eq:bnotation})
in this case arises from a three-pion one-nucleon vertex whose
explicit form can be found in Eqs.~(\ref{eq:la0}), (\ref{eq:la1}).
The  diagram II 
yields the contribution
\begin{eqnarray*}
        i M_{\text{II}} &=& \intdl \Bnotation(l,\ltild ) \ \tau_2^c \tau_2^b \ \frac{\gA}{2 \fpi^3}
        \biggl\{
            [\tau_1^a \delta^{bc} S_1 \cdot (-l+\ltild)  +  \tau_1^b \delta^{ac} S_1 \cdot (q+\ltild) +
               \tau_1^c \delta^{ab} S_1 \cdot (q-l)]  \\
            &&-\frac{1}{2\mN} i \varepsilon^{abc}  [v \cdot q S_1 \cdot (-l-\ltild)  -  v \cdot l S_1 \cdot (\ltild-q)  +
                 v \cdot \ltild S_1 \cdot (q+l)]  \\
            &&-\frac{1}{2\mN} S_1 \cdot (\pone+\ponepr)
                    [\tau_1^a \delta^{bc} v \cdot (-l+\ltild)  +  \tau_1^b \delta^{ac} v \cdot (q+\ltild) +
                     \tau_1^c \delta^{ab} v \cdot (q-l)]
        \biggr\},
\end{eqnarray*}
where $\Bnotation(l,\ltild )$ is defined in Eq.~(\ref{eq:defineB}).
Contracting isospin indexes and ignoring all p-wave terms ($\propto S_1\cdot q$) and
higher-order s-wave terms $\propto v\cdot q/ \mN \simeq \mpi/\mN$ we find
\begin{eqnarray*}
        i M_{\text{II}} &=&  \frac{\gA}{4 \fpi^3} \intdl \Bnotation(l,\ltild )
        \biggl\{
             - (\sdotl)     [4\taup + 4\taum -2\taux]
           + (\sdotltild) [ 4\taup + 4\taum +2\taux] \\
         && + \left[ - \sdotlmn     \vdotltild
           + \sdotltildmn \vdotl \right]    ( 2\taup - 2\taum)   \\
         && + \sdotpppprmn [\vdotl (4\taup + 4\taum -2\taux) - \vdotltild (4\taup + 4\taum +2\taux) ]
        \biggr\},
\end{eqnarray*}
where the integrand in the first line starts to contribute at NLO
while that in the last two lines gives  \NNLO{}  contribution.

\subsection{Diagram IIIa}
The crossed box diagram, Type IIIa, shown in Fig.~\ref{fig:allLoops}
has a more complicated structure.
In this diagram the operator $A^{abc}_{g_A^3}$ consists of a $\pi N \to \pi N$ scattering vertex, a
nucleon propagator and a $\pi NN$-vertex. Again we only need to
include the contributions from
the leading and subleading chiral Lagrangian in the vertices. We also include the
nucleon recoil correction in the nucleon propagator.
This diagram gives the following expression
\begin{eqnarray}
        i M_{\text{IIIa}} &=& \intdl \Bnotation(l,\ltild ) \ \tau_2^c \tau_2^b
                \ \frac{1}{4 \fpi^2} \varepsilon^{bad} \tau_1^d
        \left( v \cdot (l+q) - \frac{(\ponevec + \ponevecpr -\ltildvec) \cdot
                (\lvec + \qvec)}{2\mN} \right) \nonumber \\
        && \times \frac{i}{\ponezero - \ltildzero - \frac{(\ponevec-\ltildvec)^2}{2\mN} + i\nolik}
        \frac{\gA}{\fpi} \tau_1^c \left( S_1 \cdot \ltild -
               \frac{S_1\cdot (2\pone-\ltild) \vdotltild}{2\mN}\right).
\label{eq:MIIIa}
\end{eqnarray}
We have ignored here the subleading $c_i$-contributions to the
$\pi N \to \pi N$ rescattering vertex since they
are suppressed  in the momentum counting scheme due to the negligible
kinetic energy of the outgoing pion with  $q \simeq (\mpi, \vec{0})$.
We will rewrite the $\pi N \to \pi N$ vertex expression
in the integrand above 
in a way similar to the rearrangement in  Eq.~(\ref{eq:pipivert}) 
\begin{eqnarray}
    && v \cdot (l+q) - \frac{(\ponevec + \ponevecpr -\ltildvec)
         \cdot (\lvec + \qvec)}{2\mN} = \nonumber \\
    && = - \left(  \ponezero - \ltildzero - \frac{(\ponevec-\ltildvec)^2}{2\mN} \right) +
         2 \qzero - \frac{2\qvec \cdot (\ponevec+\ponevecpr-\ltildvec)}{2\mN},
\label{eq:trickIIIa}
\end{eqnarray}
where we used that  $v\cdot p_1^\prime = \ponezeropr  \simeq \ponevecpr{}^{2} / 2 \mN$.
The first term on the second line of Eq.~(\ref{eq:trickIIIa}) is identical to
the nucleon propagator in Eq.~(\ref{eq:MIIIa}) and will give a factor of $-1$
when inserted into Eq.~(\ref{eq:MIIIa}).
This factor of $-1$ together with the lowest-order contribution of the $\pi NN$-vertex,
$S_1\cdot \ltild$, give
the NLO contribution of diagram IIIa. The last term in the second line of
Eq.~(\ref{eq:trickIIIa}) contribute  to an outgoing p-wave pion and is ignored in this paper.
The $2q_0$ term in Eq.~(\ref{eq:trickIIIa}) will contribute to the \NNLO{} amplitude.
We next use the relation $ \ltild =l+\pone-\ponepr -q$ in the $\pi NN$-vertex
and in the nucleon propagator. We
ignore $p^\prime_{10} \sim q_0 \sim \mpi$ contributions and the 
recoil correction in  the propagator which are of a higher  order.
Carrying out the isospin algebra we get:
\begin{eqnarray*}
        i M_{\text{IIIa}} &= & \frac{\gA}{4 \fpi^3} \intdl \Bnotation(l,\ltild )
            \biggl\{
                (\sdotltild) + \sdotlmn (\vdotltild ) \\
            && -  \sdotpppprmn (\vdotltild )
              - \left( \frac{2 \vdotq} {-\vdotl + i\nolik} \right)
                 \left( S_1 \cdot \ltild \right)
            \biggr\} [2\taup - \taux]
\end{eqnarray*}
The first term in the curly bracket starts to contribute at   NLO. The remaining three terms
contribute to  \NNLO{}.

\subsection{Diagram IIIb}
Diagram IIIb (Fig.~\ref{fig:allLoops}) has a structure similar to diagram IIIa.
We proceed along the same lines as for the two previous diagrams and
obtain the following contribution:
\begin{eqnarray*}
        i M_{\text{IIIb}} &=& \intdl \Bnotation(l,\ltild ) \ \tau_2^c \tau_2^b (-1) \frac{\gA}{\fpi} \tau_1^b
        \left( S_1 \cdot l - \frac{S_1\cdot (2\ponepr-l) \vdotl}{2\mN}\right) \\
        && \times \frac{i}{\ponezeropr - \lzero - \frac{(\ponevecpr-\lvec)^2}{2\mN} + i\nolik}
        \frac{1}{4 \fpi^2} \varepsilon^{cad} \tau_1^d
        \left( v \cdot (-\ltild+q) - \frac{(\ponevec + \ponevecpr - \lvec) \cdot
             (-\ltildvec + \qvec)}{2\mN} \right)
\end{eqnarray*}
Using the on-shell condition for the incoming nucleon with  $ \ponezero = \ponevec{}^{\!\!2} / 2 \mN$,
we rewrite the
$\pi N \to \pi N$ vertex in a way similar to what was done for diagram IIIa
\begin{eqnarray}
    &&v \cdot (-\ltild+q) - \frac{(\ponevec + \ponevecpr - \lvec) \cdot
           (-\ltildvec + \qvec)}{2\mN} = \nonumber \\
    &&=  \left( \ponezeropr - \lzero - \frac{(\ponevecpr-\lvec)^2}{2\mN} \right) +
        2 \qzero - \frac{2\qvec \cdot
          (\ponevec+\ponevecpr-\lvec)}{2\mN} \,.
\label{eq:trickIIIb}
\end{eqnarray}
Using the relation $2q_0 = 2v\cdot q$ and keeping only terms
appropriate at the order we are working
we obtain:
\begin{eqnarray*}
        i M_{\text{IIIb}} &= & \frac{\gA}{4 \fpi^3} \intdl \Bnotation(l,\ltild )
            \biggl\{
                (\sdotl) + \sdotltildmn   (\vdotl )  \\
               && - \sdotpppprmn (\vdotl )
             + \left( \frac{2 \vdotq }{-\vdotl + i\nolik}\right)
                \left( S_1 \cdot l \right)
            \biggr\} [-2\taup -\taux]
\end{eqnarray*}
The first term in the curly bracket starts to contribute at NLO. The remaining three terms
contribute to  \NNLO{}.

\subsection{Diagram IV}
Diagram IV (Fig.~\ref{fig:allLoops}) has an operator $A^{abc}_{g_A^3}$ containing a
four-pion vertex, a
pion propagator and one $\pi NN$-vertex.
We keep the leading and next-to-leading order in the $\pi NN$-vertex and obtain
\begin{eqnarray*}
        i M_{\text{IV}} &=& \intdl \Bnotation(l,\ltild ) \  \tau_2^c \tau_2^b
            \left(\frac{\gA}{\fpi}\right) \tau_1^d
        \left( S_1 \cdot k_1 - \frac{S_1 \cdot (\pone+\ponepr) \vdotk}{2\mN} \right)
        \frac{i}{k_1^2-\mpi^2+i\nolik}  \\
        && \times
        \frac{i}{\fpi^2} \left\{
          \delta^{ab} \delta^{cd} \left[ (l-q)^2-\mpi^2 \right] +
          \delta^{ac} \delta^{bd} \left[ (\ltild+q)^2-\mpi^2 \right] +
          \delta^{ad} \delta^{bc} \left[ (k_1-q)^2-\mpi^2 \right]
        \right\}.
\end{eqnarray*}
The four-pion vertex is rewritten as a sum of six terms
\begin{eqnarray}
    &&\frac{i}{\fpi^2} \left\{
    \delta^{ab} \delta^{cd} \left[ (l-q)^2-\mpi^2 \right] +
    \delta^{ac} \delta^{bd} \left[ (\ltild+q)^2-\mpi^2 \right] +
    \delta^{ad} \delta^{bc} \left[ (k_1-q)^2-\mpi^2 \right]
    \right\} = \nonumber \\
    && = \frac{i}{\fpi^2} \biggl\{
    \delta^{ab} \delta^{cd} \left[ l^2-\mpi^2 \right] +
    \delta^{ac} \delta^{bd} \left[ \ltild^2-\mpi^2 \right] +
    \delta^{ad} \delta^{bc} \left[ k_1^2-\mpi^2 \right] + \nonumber \\
    && + \delta^{ab} \delta^{cd} \left[ -2 l \cdot q + q^2 \right] +
    \delta^{ac} \delta^{bd} \left[ 2 \ltild \cdot q + q^2 \right] +
    \delta^{ad} \delta^{bc} \left[ -2 k_1 \cdot q + q^2 \right]
    \biggr\}.
\label{eq:trick4pions}
\end{eqnarray}
The contributions from the first two terms on the r.h.s.~of Eq.~(\ref{eq:trick4pions})
are of a higher order. The reason is that each term cancels a
corresponding pion propagator
in the operator $\Bnotation(l,\ltild )$.
When one pion propagator in $\Bnotation(l,\ltild )$  is eliminated, the
large momentum, like $\vec{k}_1$ or $\vec{p}_1$, of this reaction
is no longer part of the loop integral
which, consequently, only contributes at a higher order
than what is considered in this paper. Keep in mind that $v\cdot k_1$,
$v\cdot p_1$ and $v\cdot p_2$ are all of the order $\mpi$,  whereas,
$|\vec{k}_1| \sim p= \sqrt{\mpi \mN}$.
The third term cancels the pion propagator $k_1^2-\mpi^2+i\nolik$ and
will contribute at \NLO{} and higher order in our counting.
The last three terms in Eq.~(\ref{eq:trick4pions}) start contributing from \NNLO{}.

Using $ k_1 = \ltild - l + q$, dropping terms contributing to outgoing
p-wave pions and carrying out the spin and isospin algebra  we find:
\begin{eqnarray*}
        i M_{\text{IV}} &=& \frac{\gA}{4 \fpi^3} \intdl \Bnotation(l,\ltild )
            \biggl\{
                \left[ ( \sdotl   ) 
               -  ( \sdotltild ) 
             + \sdotpppprmn \left(- \vdotl   + \vdotltild \right) \right] 6(\taup + \taum )  \\
            && + \left( S_1 \cdot k_1 \right) \biggl[
                  \frac{ (l+\ltild) \cdot q }
                {k_1^2-\mpi^2+i\nolik} \biggr] (- 8 \taux)
            \biggr\}.
\end{eqnarray*}
The first two terms in the first square bracket are \NLO{} contributions.
The remaining two terms are \NNLO{} terms.


\subsection{Box diagram a }
In the expression for the Box a diagram (Fig.~\ref{fig:allLoops}) we again
rewrite the pion-nucleon rescattering vertex as a sum of two terms similar to
what we did for the Type-III diagrams.
One of the new terms will
cancel nucleon propagator yielding an irreducible \NLO{} contribution.
In contrast to the derivation of the amplitude 
for the Type-III graphs, we here do not 
consider the contribution from the term with the (remaining)
nucleon propagator since it is reducible and thus included in the
initial $NN$ state interaction.
Using again that the sum of the two lowest orders contribute to the vertices,
we obtain from the box a diagram:
\begin{eqnarray}
        i M_{\text{Box a}} &=& \intdl \Bnotation(l,\ltild ) \  \tau_2^c \tau_2^b
\left( \frac{1}{4 \fpi^2} \right)
        \varepsilon^{cad} \tau_1^d \left( v \cdot (-\ltild+q) -
           \frac{(\ponevec + \ponevecpr + \lvec) \cdot (-\ltildvec + \qvec)}{2\mN}\right) \nonumber \\
        && \times \frac{i}{\lzero + \ponezero - \frac{(\ponevec+\lvec)^2}{2\mN}}
        (-1) \frac{\gA}{\fpi} \tau_1^b \left( \sdotl - \frac{S_1 \cdot (2\pone+l) \vdotl}{2\mN} \right).
\label{eq:Boxa}
\end{eqnarray}
To rewrite the expression in the pion-nucleon rescattering vertex we again use that
that $\ponepr$ is on-shell, i.e. $\ponezeropr   = \ponevecpr{}^{2} / 2 \mN $. The
$\pi N \to \pi N$ vertex is rewritten as
\begin{eqnarray*}
    &&v \cdot (-\ltild+q) - \frac{(\ponevec + \ponevecpr + \lvec) \cdot
(-\ltildvec + \qvec)}{2\mN} = \\
    &&=  - \left( \lzero + \ponezero - \frac{(\ponevec+\lvec)^2}{2\mN} \right) +
        2 \qzero - \frac{2\qvec \cdot (\ponevec+\ponevecpr+\lvec)}{2\mN}\,.
\end{eqnarray*}
The first term on the r.h.s. of the above expression is
identical to the nucleon propagator and will give a factor of $-1$  when inserted
into Eq.~(\ref{eq:Boxa}).
The last term is a
p-wave pion contribution and is ignored. Also the $2q_0$-term
does not need to be taken into account as it corresponds to  a reducible contribution.
Using $ 2 \pone +l = \ltild +(\pone+\ponepr)+q$, ignoring the p-wave pion terms,
and evaluating the spin and isospin structures, we find:
\[
        i M_{\text{Box a}}^\text{irred.} = \frac{\gA}{4 \fpi^3} \intdl \Bnotation(l,\ltild )
            \biggl\{
                - (\sdotl) + \sdotltildmn (\vdotl) + \sdotpppprmn (\vdotl)
            \biggr\} [2\taum +\taux]
\]

\subsection{Box diagram b  }

The Box b diagram given in Fig.~\ref{fig:allLoops} is very similar to the
Box a diagram and the evaluation procedure is similar.
We consider again only the irreducible contribution.
The diagram gives:
\begin{eqnarray}
        i M_{\text{Box b}} &=& \intdl \Bnotation(l,\ltild )  \ \tau_2^c \tau_2^b
          \left(\frac{\gA}{\fpi}\right) \tau_1^c
        \left( \sdotltild - \frac{S_1 \cdot (2\ponepr+\ltild) \vdotltild}{2\mN} \right)
     \nonumber \\
        && \times \frac{i}{ \ponezeropr + \ltildzero - \frac{(\ponevecpr+\ltildvec)^2}{2\mN}}
        \frac{1}{4 \fpi^2} \varepsilon^{bad} \tau_1^d
        \left( v \cdot (l+q) - \frac{(\ponevec + \ponevecpr + \ltildvec) \cdot
              (\lvec + \qvec)}{2\mN} \right).
\label{eq:Boxb}
\end{eqnarray}
Again, rewriting the pion-nucleon rescattering vertex using that $\pone$ is on shell,
$\ponezero   = \ponevec{}^{\!\!2} / 2 \mN$ leads to :
\begin{eqnarray}
    &&v \cdot (l+q) - \frac{(\ponevec + \ponevecpr + \ltildvec) \cdot (\lvec + \qvec)}{2\mN}
      = \nonumber \\
    &&=  \left(  \ponezeropr + \ltildzero - \frac{(\ponevecpr+\ltildvec)^2}{2\mN} \right) +
    2 \qzero - \frac{2\qvec \cdot
      (\ponevec+\ponevecpr+\ltildvec)}{2\mN} \,.
\label{eq:trickBoxb}
\end{eqnarray}
The first factor on the r.h.s.~of the Eq.~(\ref{eq:trickBoxb}),
when coupled with the nucleon propagator
in Eq.~(\ref{eq:Boxb}),
yields a factor of $1$ while the $2q_0$-term in Eq.~(\ref{eq:trickBoxb})
produces a reducible contribution included in the final $NN$ state interaction.
Using the relation
$ 2 \ponepr +\ltild = l +(\pone+\ponepr)-q$, ignoring terms leading to
outgoing p-wave pions and carrying out the spin and isospin algebra
leads to:
\[
        i M_{\text{Box b}}^\text{irred.} = \frac{\gA}{4 \fpi^3} \intdl \Bnotation(l,\ltild)
            \biggl\{
                (\sdotltild) - \sdotlmn (\vdotltild ) -  \sdotpppprmn (\vdotltild )
            \biggr\} [2\taum - \taux].
\]
Like the final expression for the  Box a diagram, the first term starts at NLO and
the next two terms are the \NNLO{} contributions to the amplitude.

\section{The evaluation of the individual $g_A$-diagrams }
\label{sec:appga1}
In this appendix we derive the expression for two-pion exchange diagram
linear in $g_A$ for s-wave pions produced.
The final expressions for the diagrams contain  \NNLO{} contributions.
The kinematics is defined in Fig.~ \ref{fig:ga1struct}.

\subsection{The Football diagram}
\label{sec:foot}

The two pion propagators are tied together in
pion-nucleon scattering vertices at both nucleons.
Since this loop diagram involve just pion propagators, we have an extra
symmetry factor $1/2$ associated with the boson loop. The football diagram shown in 
Fig.~\ref{fig:allLoops} gives the following expression:
\begin{eqnarray*}
        i M_{\text{F}} &= & \frac12 \intdl \Dnotation(l,\ltild) \
              \varepsilon^{cby} \tau_2^y \left(\frac{\gA}{2 \fpi^3}\right) \\
        && \times \biggl\{
            [\tau_1^a \delta^{bc} S_1 \cdot (-l+\ltild)  +  \tau_1^b \delta^{ac} S_1 \cdot
                (q+\ltild) +  \tau_1^c \delta^{ab} S_1 \cdot (q-l)]  \\
            &&-\frac{1}{2\mN} i \varepsilon^{abc}  [v \cdot q S_1 \cdot
                   (-l-\ltild)  -  v \cdot l S_1 \cdot (\ltild-q)  +
                   v \cdot \ltild S_1 \cdot (q+l)]  \\
            &&-\frac{1}{2\mN} S_1 \cdot (\pone+\ponepr)
                    [\tau_1^a \delta^{bc} v \cdot (-l+\ltild)  +
             \tau_1^b \delta^{ac} v \cdot (q+\ltild) +  \tau_1^c \delta^{ab} v \cdot (q-l)]
        \biggr\}.
\end{eqnarray*}
After performing some spin and isospin algebra,
dropping terms corresponding to the  outgoing p-wave pion and/or
higher-order corrections we obtain
\begin{eqnarray} \label{eq:Football} 
        i M_{\text{F}} &=& i \frac{\gA}{8 \fpi^3} \intdl \Dnotation(l,\ltild)
        \biggl\{
            \left[ (\sdotl) + (\sdotltild) \right] (-2\taux) \\
            && + \left[ \sdotlmn (\vdotltild ) - \sdotltildmn (\vdotl) \right] (2\taup - 2\taum)
        + \sdotpppprmn \left[ \vdotl  + \vdotltild \right](2\taux)
        \biggr\}.
        \nonumber
\end{eqnarray}
The obtained result requires some clarification.
Looking naively at the first line in Eq.~(\ref{eq:Football}) one may
conjecture that this diagram starts to contribute
already at NLO.  Indeed, assuming  $l_0\sim |\vec l|\sim  p$,
the dimensional analysis gives
\[
\frac{p}{ \fpi^3}\cdot \frac{1}{ \fpi^2 p^3} \cdot \frac{p^4}
{ (4\pi)^2}\sim \frac{1}{ \fpi^3}\frac{p^2}{ \mN^2}
\]
where the three terms on the l.h.s. stand for the $3\pi NN$-vertex,
the estimate of $D(l,\tilde l)$ as follows from Eq.~(\ref{eq:defineD})
and the integral measure, in order.
Above we also used that $(4\pi \fpi)^2 \simeq \mN^2$.
On the other hand, a more careful analysis shows that
the first line of the integral in Eq.~(\ref{eq:Football}),
which appears at NLO, is
\[
        i M_{\text{F}} = \taux\ \frac{\gA}{16 \fpi^5} \intdl \frac{2l_0}{(l_0^2-\vec l^2+i0)(l_0^2-\vec {\tilde l}^2+i0)}
             \left[ (\sdotl) + (\sdotltild) \right],
\]
where we have used that $l_0\sim |\vec l|\sim  p \gg \mpi$
to drop all subleading contributions including the $1/\mN$ terms in the
curly bracket of the integrand.
This last integral,
however,  vanishes when integrating over $l_0$
because the numerator of the integrand is an odd function of $l_0$
whereas the denominator is an even one.
The next-higher order contributions in Eq.~(\ref{eq:Football})
do not vanish.  They  scale as $\mpi/p$ and $p/\mN$
compared to NLO and thus emerge at \NNLO{}.
Following the same lines,  one can show that also
the other diagrams of $g_A$-topology start to contribute at \NNLO{}.

\subsection{Diagram Ia}

The double-scattering diagram Ia shown in Fig.~\ref{fig:allLoops} gives the following expression
\begin{eqnarray*}
        i M_{\text{Ia}} &=& \intdl \Dnotation(l,\ltild) \varepsilon^{cby} \tau_2^y
        \left(\frac{1}{4 \fpi^2}\right) \varepsilon^{bad} \tau_1^d
        \left( v \cdot (l+q) - \frac{(\ponevec + \ponevecpr -\ltildvec) \cdot
           (\lvec + \qvec)}{2\mN} \right)  \\
        && \times \frac{i}{\ponezero - \ltildzero - \frac{(\ponevec-\ltildvec)^2}{2\mN} + i\nolik}
        \left(\frac{\gA}{\fpi} \right) \tau_1^c \left( S_1 \cdot \ltild - \frac{S_1\cdot
            (2\pone-\ltild) \vdotltild}{2\mN}\right).
\end{eqnarray*}
In the $\pi N \to \pi N$ rescattering vertex (off nucleon 1) we included the
leading WT  vertex contribution together with its recoil correction. However, we dropped the
subleading $c_i$-terms in  this vertex since they are of  higher order (see the discussion
in the end of Sec.~\ref{sec:PC}).
The  $\pi N \to \pi N$ vertex expression is rewritten the same way as for diagram IIIa,
shown in Eq.~(\ref{eq:trickIIIa}).

Using that $\ponezeropr   = \ponevecpr{}^{2} / 2 \mN$, 
collecting the spin structures and performing the isospin algebra we get:
\begin{eqnarray*}
        i M_{\text{Ia}} &=& i \frac{\gA}{8 \fpi^3} \intdl \Dnotation(l,\ltild)
            \biggl\{
               - (\sdotl) - (\sdotltild)
               - \sdotlmn \left(\vdotltild \right) + \sdotltildmn (\vdotl)  \\
            && + \sdotpppprmn \left[\vdotl  + \vdotltild \right]
               + \frac{4 \vdotq}{-\vdotl + i\nolik} \left( S_1 \cdot \ltild \right)
            \biggr\} (\taup - \taum -\taux).
\end{eqnarray*}

This amplitude starts to  contribute at \NNLO{}, see discussion
in  Appendix~\ref{sec:foot} for more details.

\subsection{Diagram Ib}

The double-scattering diagram Ib in Fig.~\ref{fig:allLoops} gives an initial expression:
\begin{eqnarray*}
        i M_{\text{Ib}} &=& \intdl \Dnotation(l,\ltild) \varepsilon^{cby} \tau_2^y (-1)
            \frac{\gA}{\fpi} \tau_1^b
        \left( S_1 \cdot l - \frac{S_1\cdot (2\ponepr-l) \vdotl}{2\mN}\right)  \\
        && \times \frac{i}{\ponezeropr - \lzero - \frac{(\ponevecpr-\lvec)^2}{2\mN} + i\nolik}
        \frac{1}{4 \fpi^2} \varepsilon^{cad} \tau_1^d
        \left( v \cdot (-\ltild+q) - \frac{(\ponevec + \ponevecpr - \lvec) \cdot
            (-\ltildvec + \qvec)}{2\mN} \right).
\end{eqnarray*}
Again, the  $\pi N \to \pi N$ vertex is rewritten as sum of two terms as
for diagram IIIb, see Eq.~(\ref{eq:trickIIIb}).
We follow the simplifications discussed
for diagram IIIb, and use that $\ponezero   = \ponekin$.
Using  again $ l \to - \ltild$ etc. to simplify the integrals
containing the function $\Dnotation(l,\ltild)$,
collecting spin structures and performing the isospin algebra we get:
\begin{eqnarray*}
        i M_{\text{Ib}} &=& i \frac{\gA}{8 \fpi^3} \intdl \Dnotation(l,\ltild)
            \biggl\{
                (\sdotl) + (\sdotltild)
               - \sdotlmn \left(\vdotltild \right) + \sdotltildmn (\vdotl)   \\
            && - \sdotpppprmn \left[\vdotl  + \vdotltild \right]
               + \frac{4 \vdotq}{- \vdotl + i\nolik} \left( S_1 \cdot l \right)
            \biggr\} [\taup - \taum + \taux].
\end{eqnarray*}
This amplitude also starts to  contribute at \NNLO{}, see comment in  Appendix~\ref{sec:foot} for more details.

\subsection{The Mini-Football diagram }

The contribution of the mini-football diagram in Fig.~\ref{fig:allLoops} can be
written as
\begin{eqnarray*}
        i M_{\text{mF}} &=& \frac12 \intdl \Dnotation(l,\ltild)\ \varepsilon^{cby}
             \tau_2^y  \left(\frac{\gA}{\fpi}\right) \tau_1^d
        \left( S_1 \cdot k_1 - \frac{S_1 \cdot (\pone+\ponepr) \vdotk}{2\mN} \right)
        \frac{i}{k_1^2-\mpi^2+i\nolik}  \\
        && \times
        \frac{i}{\fpi^2} \left\{
          \delta^{ab} \delta^{cd} \left[ (l-q)^2-\mpi^2 \right] +
          \delta^{ac} \delta^{bd} \left[ (\ltild+q)^2-\mpi^2 \right] +
          \delta^{ad} \delta^{bc} \left[ (k_1-q)^2-\mpi^2 \right]
        \right\}.
\end{eqnarray*}
The four-pion vertex can be  rewritten as a sum of six terms
\begin{eqnarray*}
    &&\frac{i}{\fpi^2} \left\{
    \delta^{ab} \delta^{cd} \left[ (l-q)^2-\mpi^2 \right] +
    \delta^{ac} \delta^{bd} \left[ (\ltild+q)^2-\mpi^2 \right] +
    \delta^{ad} \delta^{bc} \left[ (k_1-q)^2-\mpi^2 \right]
    \right\} = \\
    &&= \frac{i}{\fpi^2} \biggl\{
    \delta^{ab} \delta^{cd} \left[ l^2-\mpi^2 \right] +
    \delta^{ac} \delta^{bd} \left[ \ltild^2-\mpi^2 \right] +
    \delta^{ad} \delta^{bc} \left[ k_1^2-\mpi^2 \right] \\
    &&+ \delta^{ab} \delta^{cd} \left[ -2 l \cdot q + q^2 \right] +
    \delta^{ac} \delta^{bd} \left[ 2 \ltild \cdot q + q^2 \right] +
    \delta^{ad} \delta^{bc} \left[ -2 k_1 \cdot q + q^2 \right]
    \biggr\}.
\end{eqnarray*}
Following the arguments outlined in the derivation of the contribution
from diagram IV,
see the discussion below Eq.~(\ref{eq:trick4pions}), most terms either
contribute to outgoing p-wave pions
or  higher orders in the chiral expansion.  The final result reads:
\[
        i M_{\text{mF}} = i \frac{\gA}{8 \fpi^3} \intdl \Dnotation(l,\ltild)
             \left( S_1 \cdot k_1  \right)
            \biggl\{
                 \frac{q \cdot (l + \ltild)}{k_1^2-\mpi^2+i\nolik}
            \biggr\} (8 \taux ).
\]
This amplitude starts to  contribute at \NNLO{}.

\section{Expressions for loop-integrals}
\label{sec:appintegr}

In this appendix we provide expressions for loop integrals
required to calculate the transition amplitude at \NNLO{}.
Using dimensional regularization and integration procedure
described in Appendix~E of Ref.~\cite{ParkMinRho93},
we obtained the following results:
\begin{equation}
\frac{1}{i} \! \int \! \! \! \frac{d^{4}l}{(2 \pi)^{4}}
\frac{v \cdot (l + \ltild) S_1 \cdot (l + \ltild)}
{(l^2 - \mpi^2 + i\nolik)  (\ltild^2 - \mpi^2 + i\nolik)  (-\vdotl + i\nolik)} \simeq 0,
\end{equation}

\begin{equation}
\frac{1}{i} \! \int \! \! \! \frac{d^{4}l}{(2 \pi)^{4}}
\frac{v \cdot (l + \ltild)}
{(l^2 - \mpi^2 + i\nolik)  (\ltild^2 - \mpi^2 + i\nolik)  (-\vdotl + i\nolik)} \simeq
-2 J(k_1^2),
\end{equation}

\begin{equation}
\frac{1}{i} \! \int \! \! \! \frac{d^{4}l}{(2 \pi)^{4}}
\frac{v \cdot (l + \ltild) \, q \cdot (l + \ltild)}
{(l^2 - \mpi^2 + i\nolik)  (\ltild^2 - \mpi^2 + i\nolik) } \simeq
2 k_1^2 (v \cdot q)  \left[ - \frac16 J(k_1^2) - \frac19 \frac{1}{(4\pi)^2} \right],
\end{equation}

\begin{equation}
\frac{1}{i} \! \int \! \! \! \frac{d^{4}l}{(2 \pi)^{4}}
\frac{(S_2 \cdot  \ltild) \, (S_2 \cdot l) \,  S_1 \cdot (l + \ltild)}
{(l^2 - \mpi^2 + i\nolik)  (\ltild^2 - \mpi^2 + i\nolik)  (-\vdotl + i\nolik)^2} \simeq
- i \varepsilon^{\mu\nu\alpha\beta} k_{1\mu} S_{1\nu} v_\alpha S_{2\beta} J(k_1^2),
\end{equation}

\begin{equation}
\frac{1}{i} \! \int \! \! \! \frac{d^{4}l}{(2 \pi)^{4}}
\frac{(S_2 \cdot  \ltild) \, (S_2 \cdot l)}
{(l^2 - \mpi^2 + i\nolik)  (\ltild^2 - \mpi^2 + i\nolik)  (-\vdotl + i\nolik)^2} \simeq
\frac34  J(k_1^2) - \frac{1}{(4\pi)^2},
\end{equation}

\begin{equation}
\frac{1}{i} \! \int \! \! \! \frac{d^{4}l}{(2 \pi)^{4}}
\frac{(S_2 \cdot  \ltild) \, (S_2 \cdot l) \,  q \cdot (l + \ltild)}
{(l^2 - \mpi^2 + i\nolik)  (\ltild^2 - \mpi^2 + i\nolik)  (-\vdotl + i\nolik)} \simeq
v \cdot q \frac{k_1^2}{2} \left[
- \frac{5}{12}  J(k_1^2) + \frac{1}{18} \frac{1}{(4\pi)^2}
\right],
\end{equation}
where integral $J(k_1^2)$ is given by Eq.~(\ref{eq:Ipi}),
and only the leading loop contributions for the s-wave pion in MCS are kept.


\end{document}